\documentclass[twoside,twocolumn,9pt]{article}
\usepackage{extsizes}
\usepackage[super,sort&compress,comma]{natbib} 
\usepackage[version=3]{mhchem}
\usepackage[left=1.5cm, right=1.5cm, top=1.785cm, bottom=2.0cm]{geometry}
\usepackage{balance}
\usepackage{mathptmx}
\usepackage{sectsty}
\usepackage{graphicx} 
\usepackage{lastpage}
\usepackage[format=plain,justification=justified,singlelinecheck=false,font={stretch=1.125,small,sf},labelfont=bf,labelsep=space]{caption}
\usepackage{float}
\usepackage{fancyhdr}
\usepackage{fnpos}
\usepackage[english]{babel}
\addto{\captionsenglish}{%
  
}
\usepackage{array}
\usepackage{droidsans}
\usepackage{charter}
\usepackage[T1]{fontenc}
\usepackage[usenames,dvipsnames]{xcolor}
\usepackage{setspace}
\usepackage[compact]{titlesec}
\usepackage{hyperref}

\usepackage{epstopdf}

\definecolor{cream}{RGB}{222,217,201}

\begin{document}

\pagestyle{fancy}
\thispagestyle{plain}
\fancypagestyle{plain}{
\renewcommand{\headrulewidth}{0pt}
}

\makeFNbottom
\makeatletter
\renewcommand\LARGE{\@setfontsize\LARGE{15pt}{17}}
\renewcommand\Large{\@setfontsize\Large{12pt}{14}}
\renewcommand\large{\@setfontsize\large{10pt}{12}}
\renewcommand\footnotesize{\@setfontsize\footnotesize{7pt}{10}}
\makeatother

\renewcommand{\thefootnote}{\fnsymbol{footnote}}
\renewcommand\footnoterule{\vspace*{1pt}%
\color{black}\vspace*{5pt}} 
\setcounter{secnumdepth}{5}

\makeatletter 
\renewcommand\@biblabel[1]{#1}            
\renewcommand\@makefntext[1]%
{\noindent\makebox[0pt][r]{\@thefnmark\,}#1}
\makeatother 
\renewcommand{\figurename}{\small{Fig.}~}
\sectionfont{\sffamily\Large}
\subsectionfont{\normalsize}
\subsubsectionfont{\bf}
\setstretch{1.125} 
\setlength{\skip\footins}{0.8cm}
\setlength{\footnotesep}{0.25cm}
\setlength{\jot}{10pt}
\titlespacing*{\section}{0pt}{4pt}{4pt}
\titlespacing*{\subsection}{0pt}{15pt}{1pt}

\fancyfoot{}
\fancyfoot[RO]{\footnotesize{\sffamily{1--\pageref{LastPage} ~\textbar  \hspace{2pt}\thepage}}}
\fancyfoot[LE]{\footnotesize{\sffamily{\thepage~\textbar\hspace{4.65cm} 1--\pageref{LastPage}}}}
\fancyhead{}
\renewcommand{\headrulewidth}{0pt} 
\renewcommand{\footrulewidth}{0pt}
\setlength{\arrayrulewidth}{1pt}
\setlength{\columnsep}{6.5mm}
\setlength\bibsep{1pt}

\twocolumn[
\vspace{1em}
\sffamily

  \noindent\LARGE{\textbf{Quantum hardware calculations of the activation and dissociation of nitrogen on iron clusters and surfaces$^\dag$}} \\
 
  \noindent\large{Georgia Christopoulou,$^{\ast}$\textit{$^{a}$} Cono Di Paola,\textit{$^{a}$}  Floris Eelke Elzinga,\textit{$^{b}$} Aurelie Jallat,\textit{$^{b}$} David Mu\~{n}oz Ramo\textit{$^{a}$} and Michal Krompiec\textit{$^{a}$}} \\

  \noindent\normalsize{Catalytic processes are the cornerstone of chemical industry, and catalytic conversion of nitrogen to ammonia remains one of the largest industrial processes implemented. Rational design of catalysts and catalytic reactions largely depends on approximate computational chemistry methods, such as Density Functional Theory, which, however, suffer from limited accuracy, especially for strongly-correlated materials. Rigorous ab-initio methods which account for static and dynamic electron correlation, while arbitrarily accurate for small systems, are generally too expensive to be applied to modelling of catalytic cycles, due to exponential time and space computational complexity. Recent advances in quantum computing give hope for enabling access to accurate ab-inito methods at scale. Herein, we present a prototype hybrid quantum-classical workflow for modeling chemical reactions on surfaces, applied to proof-of-concept models of activation and dissociation of nitrogen on small Fe clusters and a single-layer (221) iron surface. 
First, we determined the structures of species present in the catalytic cycle at DFT level and studied their electronic structure using CASSCF. We show that it is possible to decouple the half-filled Fe-3d band from the Fe-N and N-N bond orbitals, thereby reducing the active space significantly. Subsequently, we translated the CASSCF wavefunctions into corresponding qubit quantum states, using the Adaptive Variational Quantum Eigensolver, and estimated their energies using a state vector simulator, H1-1E quantum emulator and (for selected systems) H1-1 quantum computer. We demonstrated that if a sufficiently small active orbital space is chosen, ground state energies obtained with classical methods and with the quantum computer are in reasonable agreement. We argue that once quantum computing methods are scaled up so that larger active spaces are accessible, they can offer a tremendous practical advantage to the computational catalysis community. 
} \\

\vspace{0.6cm}

  ]
  
\renewcommand*\rmdefault{bch}\normalfont\upshape
\rmfamily
\section*{}
\vspace{-1cm}


\footnotetext{\textit{$^{a}$~Quantinuum, Terrington House, 13-15 Hills Road, CB2 1NL, Cambridge, United Kingdom.}}
\footnotetext{\textit{$^{b}$~Equinor ASA, Martin Linges vei 33, Fornebu, Norway.}}

\footnotetext{\dag~Electronic Supplementary Information (ESI) available. See DOI:}




\section{Introduction}
Quantum computing and quantum simulations are creating transformative possibilities by exploiting the principles of quantum mechanics in new ways to process and generate information. Although classical computational chemistry has shown great progress in predicting and describing the properties of a wide range of systems, simulating some chemical systems is still classically difficult. Therefore, there is a great interest in applying quantum algorithms to solve these problems efficiently, especially for the so-called strongly correlated systems.\cite{Lee2022} These are systems usually described by wavefunctions with a high degree of entanglement. Strong correlation is ubiquitous in the study of chemical kinetics and catalysis at the atomistic level, as well as in the modelling of light-matter interactions, magnetic materials, novel semiconductors, Mott insulators and high-temperature superconductors.\cite{Stein2017} Today's ``workhorses" among computational methods for catalysis and kinetics, while extremely useful for many use cases, are not accurate enough to drive process or materials discovery and optimisation. Quantum computing will transform this landscape by unlocking the practical potential of accurate first-principles computational methods that currently \textemdash on classical CPUs\textemdash \ rely on Density Functional Theory (DFT) and steeply scaling wavefunction methods in estimating total energies.\cite{Grumbling2019,nielsen_chuang_2010, claudino2022basics}

Several early applications of quantum computers and quantum simulations have already been presented \cite{bharti2021noisy, mcardle2020quantum, lanyon2010towards, cao2019quantum, bauer2020quantum,von2021quantum,goings2022reliably,elfving2020will}, and the initial results are quite promising. Today's quantum computers, with their 2-digit qubit counts and ever-improving gate fidelities, are sometimes referred to as NISQ (``Noisy Intermediate-Scale Quantum'') devices.\cite{1801.00862} Their computing power can be measured in terms of quantum volume, which refers to the maximum number of operations a quantum computer can perform before the signal disappears into noise. However, the quantum hardware available today needs to be further improved for quantum technology to have a significant impact on a range of industries, including the chemical industry. We believe that sufficiently large, fault-tolerant quantum computers will surpass the capabilities of currently available classical machines to simulate large, highly correlated systems.\cite{su2021fault,ding2023even} Moreover, before such large, fault-tolerant computers are available, further improvements in algorithms are expected, which would further reduce runtimes and resource requirements. Advanced modelling capabilities, such as those offered by quantum computing, will not only provide new insights into the fundamentals but, thanks to their ever-increasing accuracy, will also accelerate progress in research by enabling fast \textit{in silico} experimental testing of new ideas, thus reducing the number of costly ``trial-and-error" experiments in the laboratory.

With this in mind, the aim of this study is to demonstrate the workflow of a typical surface science ab initio simulation on quantum computers. To this end, we focused on the activation and dissociation of nitrogen on iron clusters. Although the production of ammonia is the oldest and one of the largest catalytic processes in the chemical industry, it still requires high temperatures and pressures. Therefore, considerable efforts are being made to develop novel catalysts and methods for milder, more environmentally friendly activation of nitrogen.\cite{Schlgl2003,Tanabe2013,Wang2018,Marakatti2020} Catalytic cycles on conventional catalysts for ammonia synthesis have been rationalised via DFT models. \cite{ZeinalipourYazdi2021} However, predictive simulation of these processes requires first-principles methods that correctly describe the strongly correlated properties of the catalysts. Strong magnetic interactions, such as those found in the Haber-Bosch catalysts,  are thought to be important for their catalytic activity.\cite{Biz2021,Munarriz2018}
The FeMo cofactor, nature's solution for ammonia synthesis \cite{Einsle2014}, is beyond the reach of current computational chemistry methods \cite{Li2019} and is considered a prime target for quantum-accelerated computational chemistry.\cite{PRXQuantum.2.030305} Moreover, with reference to the currently available literature\cite{liu2018heterogeneous}, it is believed that Fe$_3$ and  Fe$_4$ clusters on the $\theta$-Al$_2$O$_3$(010) surface can be successfully used as a heterogeneous catalyst for ammonia synthesis. 

Larger model systems such as iron clusters and surfaces require quantum phase estimation (QPE) algorithms. However, their resource requirements are too large for the currently available quantum computers. Variational algorithms are considered the most suitable techniques for NISQ devices. In these algorithms, a hybrid quantum-classical setup is constructed in which a relatively flat-parameterised quantum circuit performs heavy tasks such as encoding correlated molecular wavefunctions to calculate the expected value of the energy, while the classical computer collects the quantum computer data to optimise the parameters within the variational loop. In this study, the variational quantum eigensolver (VQE) algorithm \cite{peruzzo2014variational} was used, which is one of the most commonly used variational algorithms to perform quantum chemical simulations on NISQ devices.

This work is structured as follows. In Sect. \ref{method}, the classical and quantum computational methods of the simulations are discussed. In sec. \ref{res}
we provide all the results on the iron clusters and surfaces with nitrogen, starting with classical calculations to find the best way to simplify the system for calculations with quantum algorithms. Finally, in the last section (sec. \ref{conclusions}), we present a detailed summary of our results and future work.

\section{Methods} \label{method}
To model the activation and dissociation of nitrogen on iron clusters and surfaces, we used a wide range of classical and quantum computational methods. First, we built the atomistic models and optimised them classically, and then used these data to build our quantum models. This section is structured as follows: First, we describe the computational details and classical optimisation methods used for the iron clusters and surfaces. Then we present the technical details for the quantum calculations including the Hamiltonian construction and the details of the quantum hardware used. In the last section, we mention all the equations and approaches used in the various energy calculations.

\subsection{Hartree-Fock, CASSCF and DFT calculations.}

Atomistic models of nitrogen activation on iron were built by running simulations for the bulk metal and for the single/multi-layer surface slabs using the Quantum Espresso (QE) Kohn-Sham Density Functional Theory (DFT) package.\cite{QE_09, QE_17} Spin-polarised simulations were performed using DFT-D3 van der Waals dispersion correction\cite{grimme3} to optimise the geometry of the structures of interest. In all DFT calculations, we employed the Perdew-Burke-Ernzerhof (PBE) GGA exchange-correlation functional together with projector augmented-wave (PAW) pseudopotentials.\cite{Blochl1994} Wavefunction and charge density cut-offs of 36-47 Ry and 221-448 Ry, respectively, were used together with a Brillouin zone sampling mesh of (12x12x12) for the bulk system and a Marzari-Vanderbilt smearing of 0.04 Ry. Single-point calculations were performed at the restricted open-shell Hartree-Fock (ROHF) and CASSCF levels of theory and with the Los Alamos National Laboratory
(lanl2dz) effective core potential (ECP) and the basis set in PySCF.\cite{sun2018, sun2020}

A cubic box of at least 10 \AA\ for the iron clusters and ~35 \AA\ of vacuum on top of single/multi-layer surfaces (z-axis) were adopted to avoid any density overlap along non-periodic directions. The PBE0 functional was also used here, which mixes the PBE exchange energy and the Hartree–Fock (HF) exchange energy in a 3:1 ratio together with the full PBE correlation energy. In addition to building the structures, the Nudged Elastic Band (NEB) method\cite{jonsson1998} available in the Quantum Espresso package\cite{QE_09, QE_17} was used to find transition states and minimum energy paths between known reactants and products. A loose force convergence threshold of approximately 0.05 eV/\AA~ was also used.

Periodic HF mean-field calculations were performed using the Los Alamos National Laboratory (lanl2dz) basis/pseudopotentials of the double-$\zeta$ type for both nitrogen and iron atoms, as implemented in PySCF. For the clusters, first, a ROHF calculation and then a CASSCF calculation was performed, each time using the CI coefficients and orbitals from the previous CASSCF calculation as the initial guess. To ensure convergence of the total energy, a general second-order solver called the Co-Iterative Augmented Hessian (CIAH) method, implemented in PySCF, was used. The same setup was applied to the iron surface models by sampling the Brillouin zone only at the $\Gamma$-point and using a Gaussian auxiliary basis set for density fitting in the form available in PySCF. 

\subsection{Atomic Valence Active Space (AVAS) orbital localisation.}

To construct the active orbital space for post-Hartree-Fock and quantum calculations, we used our own implementation of Regional Embedding (RE) \cite{RegionalEmbedding}, a variant of the AVAS method.\cite{AVAS} In AVAS, the active space is constructed by selecting a list of atomic orbitals (projectors) defined as spherically averaged ground state HF wavefunctions of free atoms in a minimal basis (MINAO).  In the RE variant and only for the virtual orbitals, basis functions of the current computational basis are used. An overlap matrix of the occupied orbitals is calculated, which is projected into the space of these selected atomic orbitals:
\begin{align}
[ S^{A}]_{ij} = \langle i|\hat P|j \rangle
\end{align}
Next, a matrix of eigenvectors $[ U]_{ij}$ is computed such that:
\begin{align}
 S^{A} U = U
 \operatorname{diag}(\sigma_1,\ldots,\sigma_{N_\mathrm{occ}})
\end{align}
Now there are at most as many non-zero eigenvalues as there are selected atomic orbitals. This matrix defines a rotation of the occupied orbitals, separating them into two groups: those that have a non-vanishing overlap with the target atomic orbitals ($\sigma_i\neq 0$) and the remaining ones that have exactly zero overlap with our target space. The latter can remain inactive (as nuclear orbitals), and the former are the active occupied orbitals. An analogous transformation is performed for the virtual orbitals. To further reduce the size of the active space, we remove orbitals whose overlap $\sigma_i$ is lower than a certain threshold called the AVAS Overlap Threshold.

The new set of rotated molecular orbitals in the basis of the fragment projectors defines a sorted descending set of eigenvalues for the occupied and unoccupied orbitals. These values represent the AVAS overlap threshold in the range between 1.0 (highest overlap) and 0.0 (lowest overlap).

\subsection{Technical details for the quantum calculations.}

To calculate the total single-point energies of products, transition states and reactants to evaluate the reaction energies, quantum simulators and state-of-the-art quantum hardware were used. This was achieved by using a stack of PySCF\cite{sun2018}, InQuanto\cite{inquanto} and Pytket\cite{Sivarajah_2021} as well as Qulacs \cite{suzuki2021qulacs} statevector simulator, Qiskit Aer simulator \cite{Qiskit} and the 'H1-1' quantum device and its emulator.

The second-quantized Hamiltonians for the VQE calculations were obtained by Hartree-Fock calculations in the PySCF extension of InQuanto using the lanl2dz basis set and the corresponding ECP, localising the orbitals with AVAS, and transforming the Hamiltonian to the MO representation in InQuanto. The terms with coefficients smaller than a certain threshold (i.e. below $10^{-8}$) were then removed.

For the state preparation, we applied the Adaptive Derivative-Assembled Pseudo-Trotter (ADAPT \cite{Grimsley2019}) approach, which uses the Unitary Coupled Cluster Singles Doubles formalism (UCCSD) excitation pool and the 'Chemically Aware' \cite{Khan2022} circuit synthesis method. The Jordan-Wigner transformation was used. In the final step, all circuits are optimised and compiled with pytket for the target backend. The ADAPT-VQE algorithm was run on the Qulacs statevector simulator to build the ansatz and determine its parameters. The expectation value of the energy was then evaluated using Hamiltonian averaging. We measured the Pauli terms by appending measurement circuits directly to the system register and using Partition Measurement Symmetry Verification (PMSV) error mitigation.\cite{Yamamoto2022} The N-electron Valence State 2$^{nd}$-order Perturbation Theory (NEVPT2) dynamical correlation \cite{Angeli2002,krompiec2022} was applied to the iron surface models only and evaluated using the CASSCF reduced density matrices.

The measurement circuits were then prepared, optimised and delivered for processing on quantum emulators and hardware using pytket. For our 6-qubit calculations, we used the emulation and processing of the Quantinuum H-Series, 'H1-1', 20-qubit device. The 'H1-1' device is constantly being developed and upgraded. The experiments presented in this study were conducted from March to May 2023. The device specifications can be found online.\cite{2023GitHub}

\subsection{General details of the energy calculations}

\begin{figure}
 \centering
\includegraphics[height=3cm]{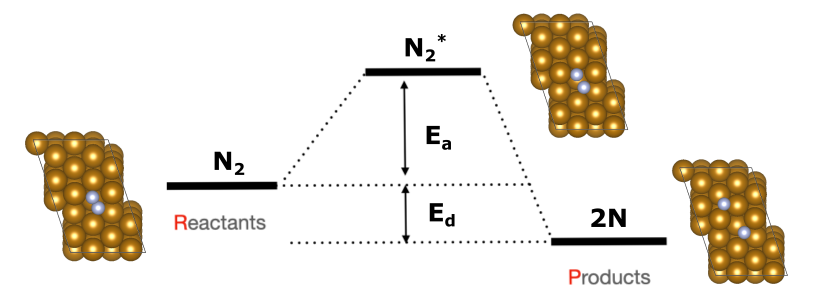}
 \caption{Generic energy profile scheme for the nitrogen dissociation step in the ammonia synthesis process. Fully relaxed structures of Fe/N$_2$ and Fe/2N are also shown. Colour code: Gold for Fe and silver for N.}
 \label{fgr:generic}
\end{figure}

The focus of this study is on the calculation of the activation and dissociation energies of nitrogen on iron clusters and surfaces. Figure~\ref{fgr:generic} shows the quantities we would like to calculate

\begin{equation}
 E_a=E_{Fe/N_2}^{tot}-E_{Fe/N_2^*}^{tot}
 \label{eq:Eact}
\end{equation}

representing the electronic activation energy with respect to the kinetic constant of the process, and

\begin{equation}
 E_d=E_{Fe/N_2}^{tot}-E_{Fe/2N}^{tot}
 \label{eq:Ediss}
\end{equation}

representing the electronic component of the thermodynamical driving force for surface-assisted molecular dissociation (i.e. neglecting the vibrational and entropy contributions). In the figure, the initial, transition and final states of the iron slab are shown for illustration. In the case of the iron clusters, the same approach was followed.

In the quantum emulator and hardware experiments, the approximate correlation energy $E_{\mathrm{corr}}$ is calculated, which is defined as the difference between post-HF and HF energies.
The $\Delta E(\boldsymbol{\theta})$ can be defined as

\begin{equation}
 \Delta E(\boldsymbol{\theta})
 =
 E_{\mathrm{total}}(\boldsymbol{\theta})
 -
 E_{\textrm{ HF }}^{\circ}
 \label{eq:delta-e}
\end{equation}

where $E_{\textrm{ HF }}^{\circ}$ denotes the
HF energy which is calculated with the classical computer and
$E_{\textrm{total}}(\boldsymbol{\theta})$
refers to the total energy calculated on quantum hardware or simulator, which may be affected by noise and/or stochastic errors.
If these errors are small enough, the value of $\Delta E(\boldsymbol{\theta})$ calculated with optimal parameters
$\boldsymbol{\theta}$ is a good approximation to
$E_{\mathrm{corr}}$,
which must be a negative value.
However, quantum noise in the NISQ device can cause $\Delta E(\boldsymbol{\theta})$ to even take on a positive value, since noise-induced high energy excited states can contaminate the calculated ground state wavefunction.

\subsection{Technical details of emulator experiments.}

Before running experiments on the quantum computer, we performed a thorough analysis and benchmarking using classical quantum simulators, to investigate the convergence with respect to the number of measurements (``shots'') and the effect of quantum errors (noise).  The quantum circuits were built using the exponents (fermionic excitations) selected by the ADAPT-VQE algorithm. The start geometry of the Fe$_3$N$_2$ cluster was used in all the benchmark simulations.

\begin{figure}
   \centering
\includegraphics[height=6cm]{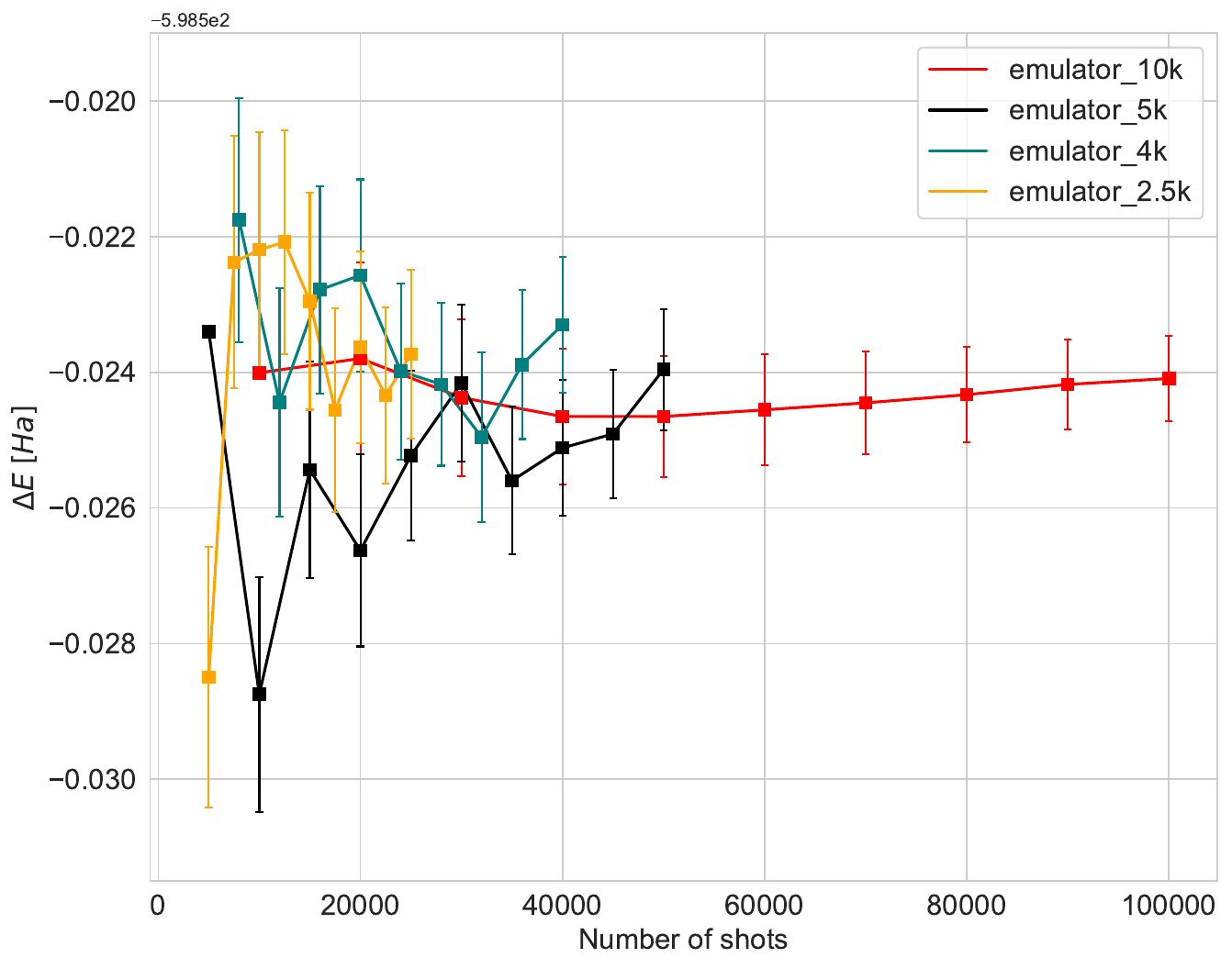}
   \caption{Expectation value obtained by changing the number of shots used in the computations with the Quantinuum 'H1-1E' noisy emulator backend. The Fe$_4$N$_2$ cluster's initial geometry was consistently utilized across all simulations. The error bars in the results represent the standard deviation.}
    \label{fig:shots}
\end{figure}

\begin{figure}
   \centering
\includegraphics[height=5cm]{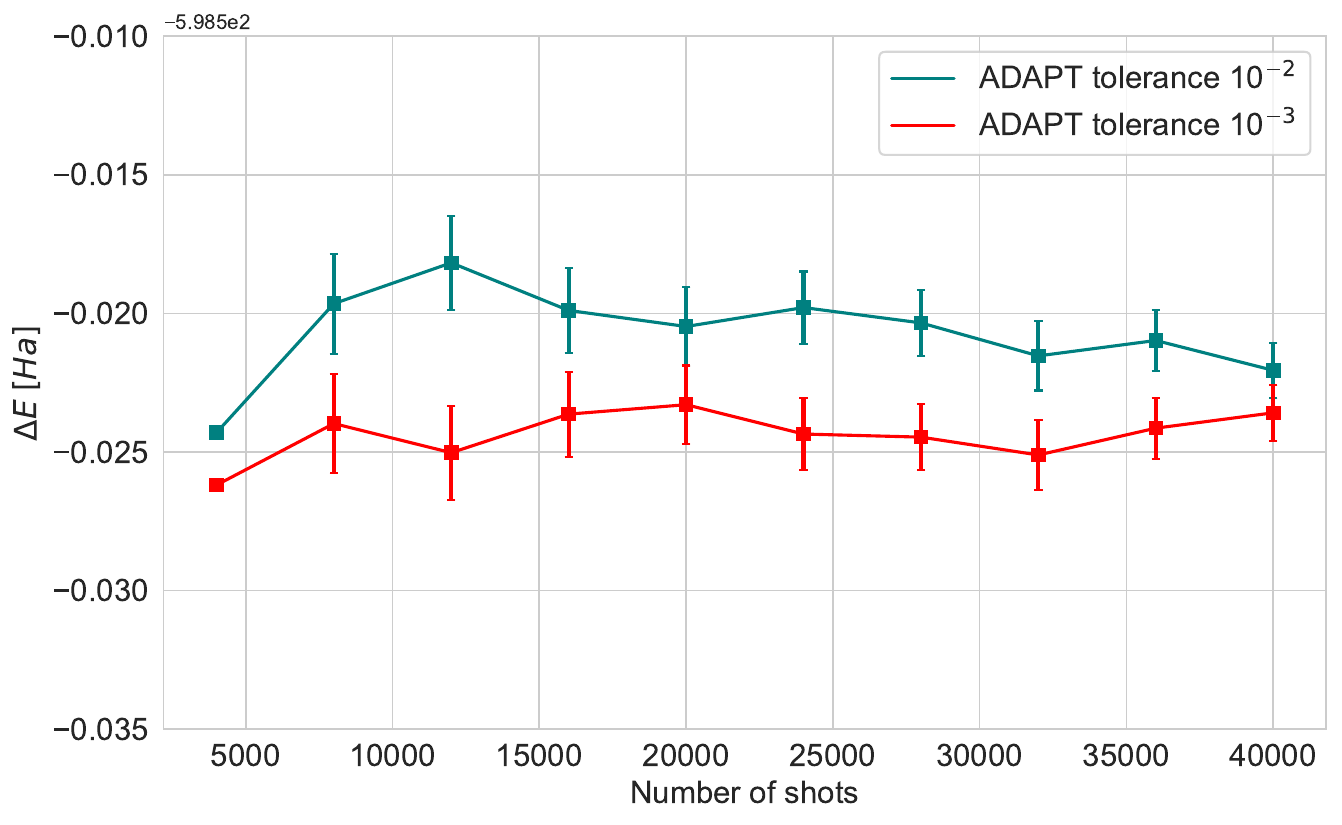}
   \caption{Expectation value obtained by varying the ADAPT-VQE threshold when running emulator ('H1-1E') experiments with 4K shots. The Fe$_4$N$_2$ cluster's initial geometry was consistently utilized across all simulations. The error bars in the results represent the standard deviation.}
   \label{fig:4k_start}
\end{figure}

Simulating an actual machine is a key step towards understanding the resources needed to run on hardware. Figure~\ref{fig:shots} shows the relationship between the number of shots and the accuracy of calculations performed using the initial geometry for the Fe$_4$N$_2$ cluster. The cost of a hardware experiment depends on the type and number of gates used for the circuits, as well as the total time required for all processes to be compiled. In figure~\ref{fig:shots}, an approach called batching has been used, where we repeat each experiment with a target number of shots ten times. In the case of 4k shots, for example, the target number is 4k, and after 10 repetitions, we have an experiment with 40k shots in total. The energy and error bars shown in  figures~\ref{fig:comparison} - ~\ref{fig:4k_start} are the mean and standard deviation, which were obtained at the end of each repetition. We can see that error bars get significantly smaller when there are enough samples. For the case of a hardware device, only a single batch of experiments will be run. To limit the effect of noise when estimating expectation values, PMSV (Partition Measurement Symmetry Verification) noise mitigation method has been employed.  This reduces experimental noise by removing shots in which a symmetry-breaking error has occurred. Here, the mirror planes ($Z_2$) and electron-number conservation ($U_1$) symmetries have been used which can be represented by a single Pauli string that tracks the parity of the wavefunction.

To reduce circuit depth and limit the magnitude of noise, we investigated the effect of relaxing the ADAPT convergence threshold from $10^{-3}$ to $10^{-2}$, see figure~\ref{fig:4k_start}. Clearly, decreasing the threshold negatively affects accuracy, but the maximum energy difference was 0.07 Ha, which we consider to be within the acceptable limits for VQE experiments.Taking into consideration the different results and the limited availability of the actual hardware device, the optimal choice was to conduct experiments with 4000 shots.

Finally, six qubits with three parameters and twenty 2-qubit gates were used in the circuits designed for the start and final geometry. The deepest circuit in this study, the TS geometry utilised six qubits as well, but this time four parameters and thirty-nine controlled-NOT gates were used. Figures ~\ref{fig:statevector} and ~\ref{fig:comparison} in  Supplementary Information\dag shows the convergence of the expectation value of energy with respect to number of measurements in the absence of noise .

\section{Results} \label{res}

\subsection{Simulations of iron clusters.}

\begin{figure}
 \centering
 \includegraphics[height=2cm]{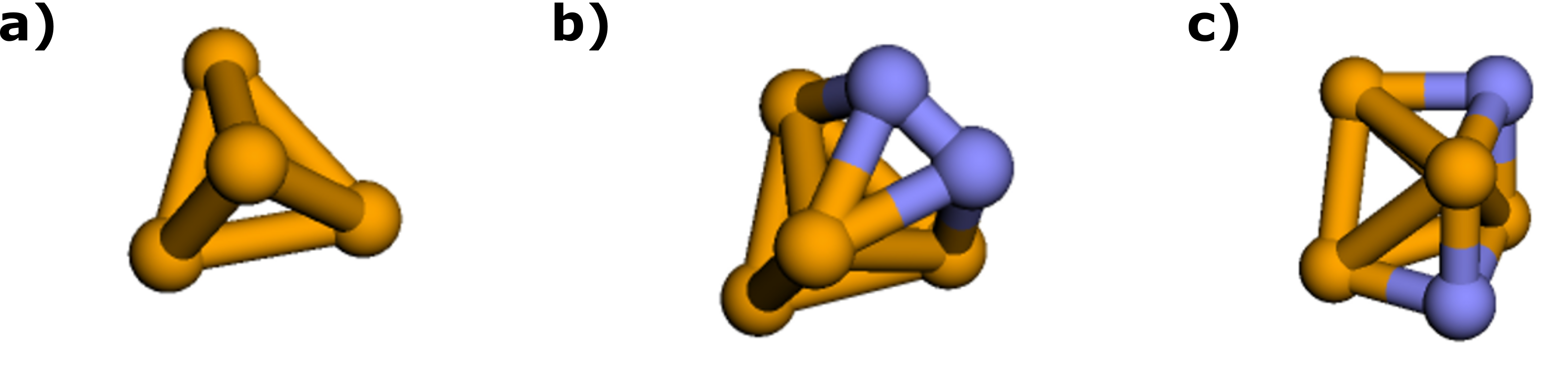}
 \caption{Visualisation of the optimised geometries of the a)Fe$_4$ cluster b) N$_2$ adsorption on Fe$_4$ and c) N$_2$ dissociation on Fe$_4$ using InQuanto-NGLView. Colour code: orange for Fe and blue for N.}
 \label{fig:clusterFe4}
\end{figure}

It is shown\cite{liu2018heterogeneous,mcwilliams2015dinitrogen,vcoric2015binding} that Fe clusters promote N$_2$ reduction. Thus, the applicability of our hybrid quantum-classical workflow was first tested on small iron clusters. For the Fe$_4$ cluster (figure~\ref{fig:clusterFe4}(a)), which is the smallest 3D iron cluster, different multiplicities of the ground state have been found in the literature.\cite{pakiari2010detailed,liu2018heterogeneous,gutsev2021evolution} Therefore, the geometry of Fe$_4$ was optimised for several possible spin multiplicities (7, 9, 11 and 13) and we found that the state with multiplicity 7 has the lowest energy. Here we focus on the catalytic behaviour of the Fe$_4$ clusters, while we also studied the Fe$_3$ cluster and its relevant products by adding two nitrogen atoms similar to the case of Fe$_4$ 
(Supplementary Information\dag). As with the Fe$_3$ cluster, after optimising the Fe$_4$ system, the Fe$_4$N$_2$ clusters were constructed by adding two nitrogen atoms and re-optimising (initial geometry, figure~\ref{fig:clusterFe4}(b)) or by stretching only the nitrogen atoms and re-optimising to obtain the final geometry (figure~\ref{fig:clusterFe4}(c)).

\begin{figure}
 \centering
 \includegraphics[height=6cm]{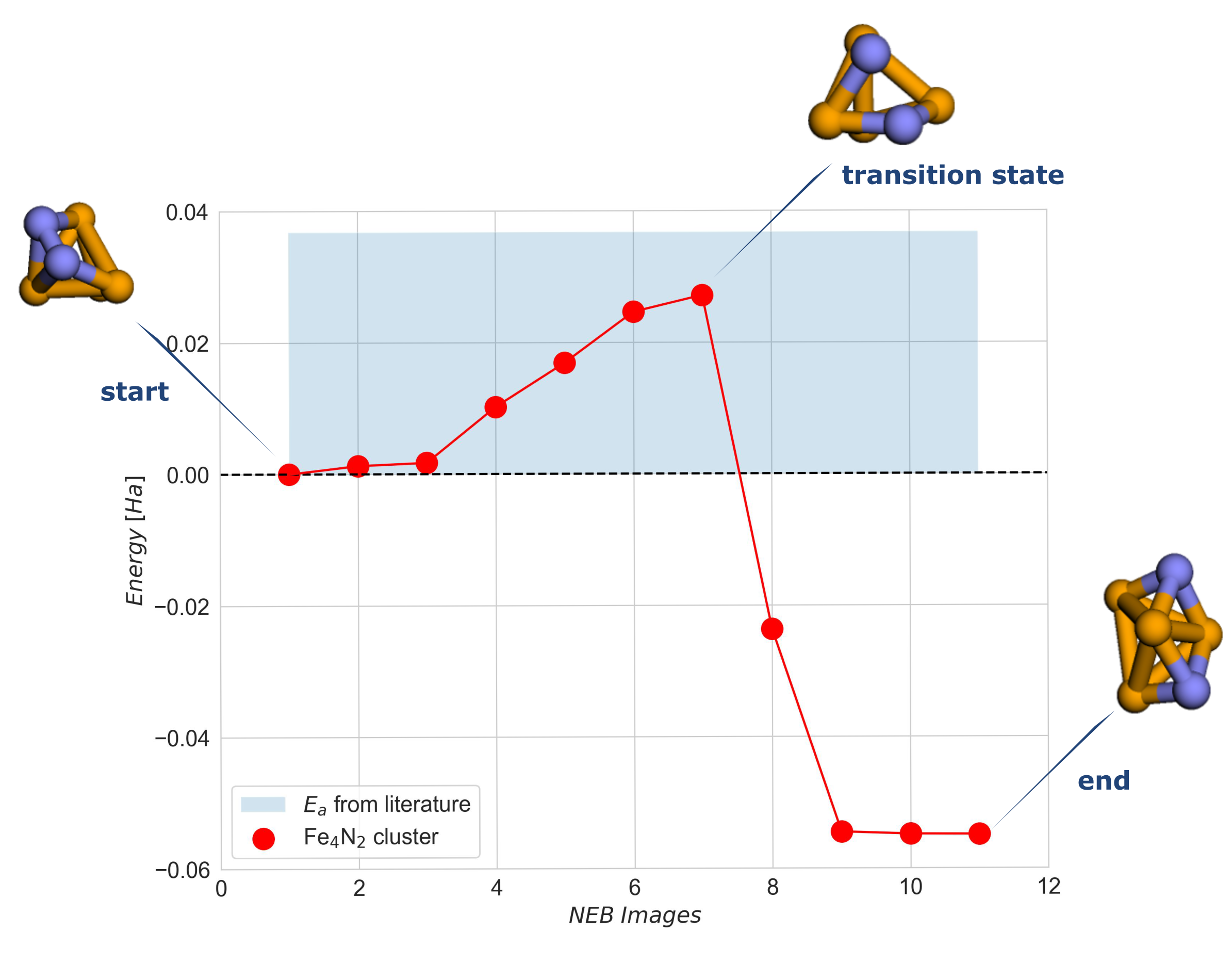}
 \caption{ NEB simulation of the minimum energy path for the dissociation of the N$_2$ molecule on the Fe$_4$N$_2$ cluster. The light blue region represents the range of activation energy values available in the literature.\cite{gutsev2021evolution,liu2018heterogeneous} Colour code: orange for Fe and blue for N.}
 \label{fig:neb_cls4}
\end{figure}

To calculate the activation energy E$_a$ we had to determine the structure of the transition state (TS), N$_2$ for the Fe$_4$ cluster. The results for Fe$_4$N$_2$ with eleven NEB images are shown in figure~\ref{fig:neb_cls4}. The activation energy calculated here is in the range of values found in the literature.\cite{gutsev2021evolution,liu2018heterogeneous}

\subsubsection{Active space selection}

Due to hardware limitations, active space approximations were employed. It is clear that the use of a small active space leads to neglection of the dynamical correlation energy, which is the dominant part of the correlation energy of these systems. However, a minimal active space should be sufficient to capture any static correlation, provided the orbitals are carefully localised, selected and optimised. To this end, we have used the AVAS (Atomic Valence Active Space) localisation procedure and the CASSCF orbital optimisation.

If one constructs the active space from (partially) filled Fe d orbitals and the N-N occupied and virtual orbitals, no excitations from Fe 3d to N-N $\sigma$ or $\pi$ orbitals are found in the CI wavefunction. We have also separately verified that excitations from the occupied N-N orbitals to the empty Fe-3d orbitals do not contribute. This means that the excitations in the Fe d manifold and in the N-N bond are not coupled and that a reduced active space can be created consisting of orbitals of the dinitrogen system and selected Fe orbitals interacting with it, i.e. excluding all half-filled 3d-like orbitals on Fe.

As can be seen in Table~\ref{tbl:fe4_NFe}, the contribution of the excited determinants in the wavefunction is at least 10$\%$ and only the first, second and third active  orbitals are significantly active. Supplementary Information\dag \ figure~\ref{fig:orbitals_Fe4N2} contains an illustration of these orbitals. For the construction of the AVAS active space, we selected 2p orbitals of the two nitrogen atoms and doubly occupied or empty 3d orbitals of the five iron atoms. The overlap threshold for occupied orbitals was 0.9, while the threshold for virtual orbitals was set at 0.97. The half-filled orbitals of Fe atoms were frozen and thus excluded from the active space.

\begin{table}[h]
\small
    \caption{Largest CI components of the transition state for the Fe$_4$N$_2$ cluster when the 2p orbitals of the two nitrogen atoms and 3d orbitals of the five irons are selected}
    \label{tbl:fe4_NFe}
    \begin{tabular*}{0.4\textwidth}{@{\extracolsep{\fill}}lll}
    \hline
    alpha occ-orbitals & beta occ-orbitals & CI coefficient \\
    \hline
    $[0 1 2]$ & $[0 1 2]$      &         0.98709  \\
    $[0 1 3]$ & $[0 1 3]$      &        -0.12552  \\
    $[0 1 3]$ & $[0 2 3]$      &         0.03952  \\
    $[0 1 3]$ & $[0 1 4]$      &        -0.04160  \\
    $[0 2 3]$ & $[0 1 3]$      &         0.03952  \\
    $[0 1 4]$ & $[0 1 3]$      &         -0.04160 \\
    \hline
  \end{tabular*}
\end{table}

\subsubsection{Energy calculations.}

Along the minimum energy path derived from the NEB calculation, energy calculations using DFT, HF, CASSCF, and classical VQE statevector were carried out. The CASSCF orbitals were used to construct the Hamiltonian for VQE, starting from the AVAS-localised orbitals (point 1) or from the orbitals of the previous data point.

\begin{figure}
    \centering
   \includegraphics[height=6cm]{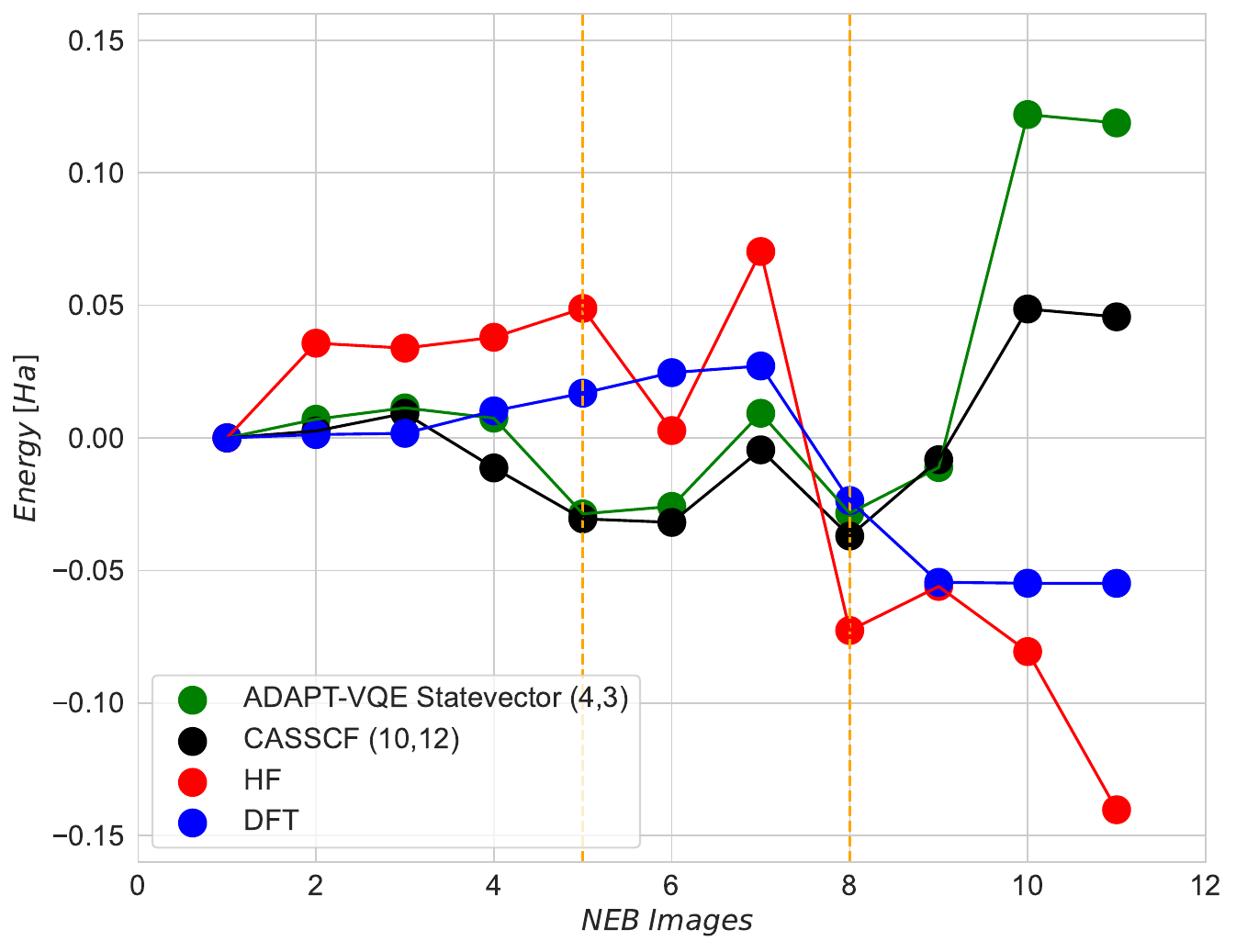}
    \caption{Relative energies (E$_{first}$-E$_{current}$) of various Fe$_4$N$_2$ geometries obtained by using a) classical ADAPT-VQE statevector with a (4,3) active space; b) CASSCF with a (10,12) active space; c) HF and ) DFT.}
    \label{fig:ene_cluster}
\end{figure}

The CASSCF and the classical VQE statevector results are quite unexpected. As can be seen in figure~\ref{fig:ene_cluster}, the energy of the final state is higher than that of the initial and transition states. Also, the trend of the CASSCF and VQE energy graphs does not follow the DFT. The most likely reason for this is that the potential energy surfaces at the CASSCF and DFT levels of theories are qualitatively different, especially at the middle and the end of the NEB path. However, the middle section of the CASSCF and VQE curves still exhibits a peak, which can be taken as an approximation to the transition state. Therefore, we decided to focus only on the range between images 5 and 8, i.e between two dashed orange lines shown in figure~\ref{fig:ene_cluster}. Thus, for the next calculations, the initial state is the 5th image, the transition state is the 7th image and the final state is the 8th image of NEB calculation. The energy difference between the zero point energy (NEB image 1) and the present NEP image is used to determine energies in figure~\ref{fig:ene_cluster}.

\subsubsection{VQE emulator and hardware experiments.}

\begin{figure}
\centering
\includegraphics[height=6cm]{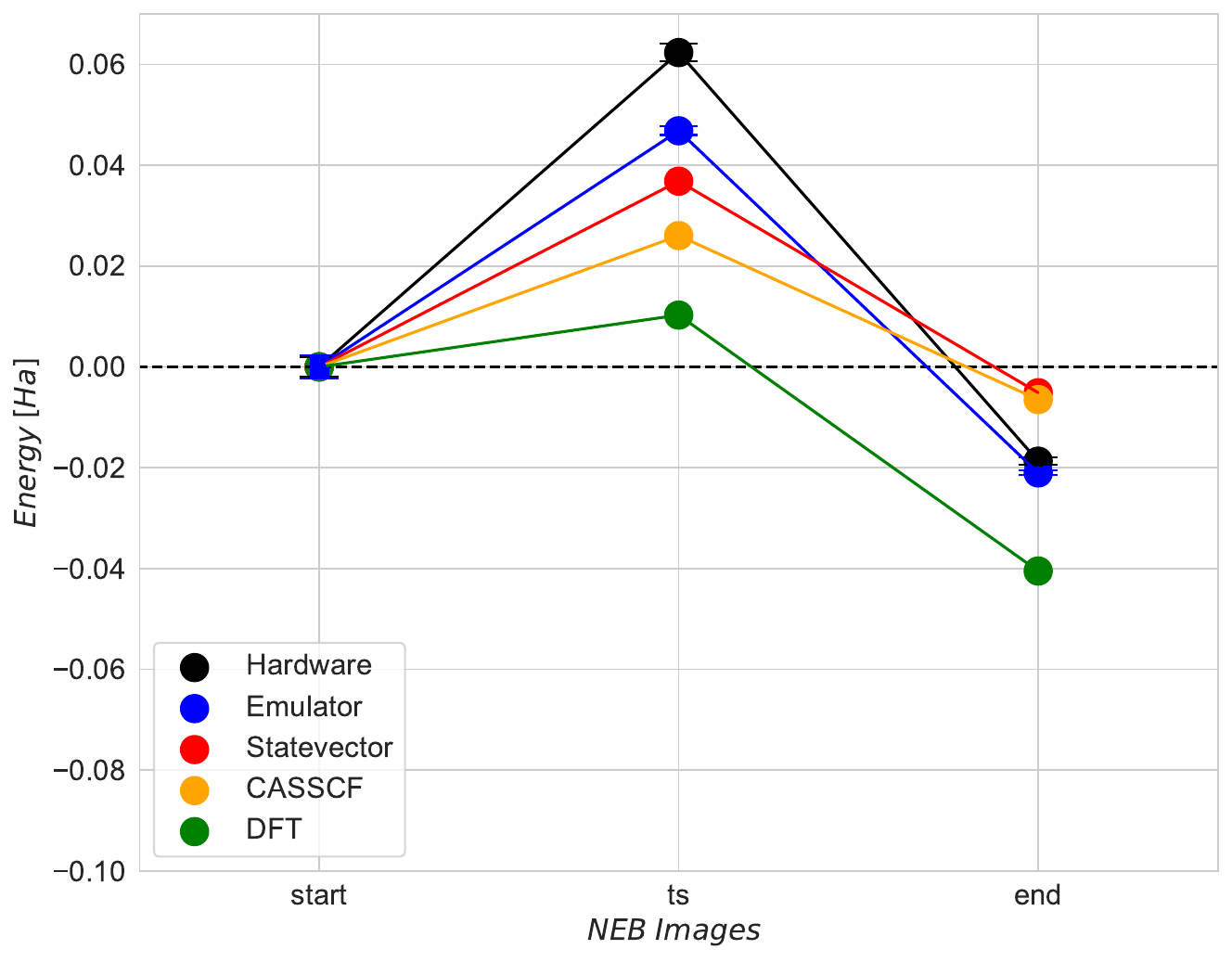}
\caption{Comparison of the activation (E$_a$) and the dissociation (E$_d$) energies by using ADAPT-VQE with a) Quantinuum 'H1-1' device, b) Quantinuum 'H1-1E' noisy emulator backend, c) Qulacs backend statevector emulator and by using d) DFT. The error bars in the hardware and emulator results represent the standard deviation.}
\label{fig:vqe_compare}
\end{figure}

In this section, we compared the Fe$_4$N$_2$ cluster activation and dissociation energies calculated using statevector, emulator, hardware VQE, and DFT. The aim was to capture correlation energy equivalent to CASSCF. Therefore, we conclude that the 'H1-1E' results are particularly reliable, as illustrated in figures~\ref{fig:ene_cluster} and ~\ref{fig:vqe_compare}, where ADAPT-VQE statevector energy values with a (4,3) active space are quite similar to those obtained from CASSCF with a (10,12) active space. Despite the iron clusters not being strongly correlated, the dissociation energies are reproduced at a reasonably good level. As depicted in figure~\ref{fig:vqe_compare}, the data points computed on the hardware are in excellent agreement with the emulation results. The activation energy calculated using emulator and hardware experiments is not close to the activation energy obtained using DFT. However, the dissociation energy only varies by less than 0.02 Ha. Moreover, the deviation between state vector simulations and quantum hardware (or noisy emulation) is much larger than the finite-sampling error (denoted by the error bars). 

Comparing the energy results computed classically with the ADAPT-VQE statevector to those from emulator and hardware experiments, we can see that the highest level of accuracy is achieved for the final state in both the emulator and hardware (Table~\ref{tbl:comp_clust}). However, it is about 7 times worse than the so-called ``chemical accuracy" target of 1 kcal/mol (1.6 mHa).

\begin{table*}
\small
  \caption{Comparison of the energy values of ADAPT-VQE when running classical statevector, emulator and hardware calculations}
  \label{tbl:comp_clust}
   \begin{tabular*}{\textwidth}{@{\extracolsep{\fill}}llllll}
    \hline
    State & State vector (Ha) & 'H1-1E' Emulator (Ha) & Deviation (mHa) & 'H1-1' Hardware (Ha) 
    & Deviation (mHa)\\
    \hline
    Initial & -598.561 & -598.523 & 38 & -598.524 &  0.74\\
    Transition &-598.523 & -598.476 & 47 & -598.461  & 14\\
    Final & -598.556 & -598.544 & 12 & -598.542  & 1.6 \\
    \hline
  \end{tabular*}
\end{table*}

\subsection{Simulations of iron surfaces.}

The high quality of the experimental output obtained on the Quantinuum 'H1-1' device (and 'H1-1E' noisy emulator) has shown encouraging results, indicating that this workflow is suitable for larger simulations. Thus, our next target was to study the activation and dissociation of nitrogen on iron surfaces. 

To model the dissociation on the Fe catalyst, the equilibrium value of the lattice parameter was first calculated. A final value of 2.829 \AA \ was found, which is very close to the lattice parameter available in the literature of about 2.832 \AA. ~\cite{zhang2019interplay} Then, the orientation of the simulated surface was determined. It has been shown~\cite{zhang2019interplay} by comparing the relative difference in intrinsic reactivity of various iron facets that the reaction rate, relative to the (100) facet follows the order of (221) $>$ (311) $>$ (111) $>$ (211) $>$ (310) $>$ (210) $>$ (110), with 5.59, 5.13, 4.92, 4.32, 4.19, 2.40, and 1.90 orders of magnitude higher than the (100) facet, respectively. Thus, we opted for the (221) direction, even though this was not the simplest case. We constructed the atomistic structure by selecting the number of repeated unit cells along the \textit{X} and \textit{Y} axes of the slab, determining the number of layers, and specifying the vacuum thickness above and below the structure (along the non-periodic 'z' axis).

We found that a (3x3x1) monolayer slab with a total amount of about 20 \AA\ of vacuum, equally distributed around the model, was large enough to prevent density overlapping between neighbouring cells along the \textit{Z} axis and ensure enough surface space for the adsorption of a single molecule. The slab was generated using the model-building functions of the Atomic Simulation Environment (ASE).\cite{ask2017} Even though we can consider the 3x3x1 slab good enough for testing purposes, more ``realistic" models should include at least 7 layers to mimic the bulk behaviour in the innermost region, as suggested by Kaushal et al. \cite{Kaushal2011} and Krupski et al. \cite{Krupski2015}

\begin{figure}
\centering
\includegraphics[height=3cm]{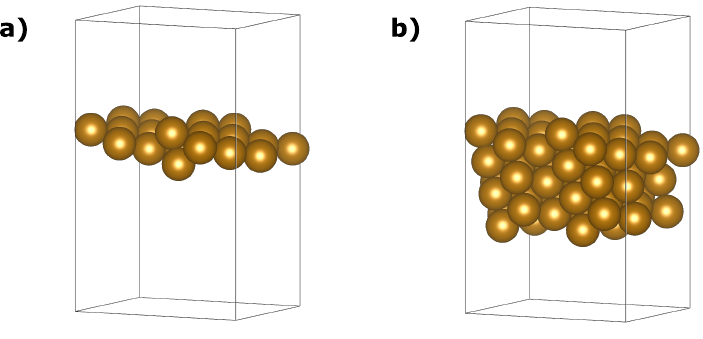}
\caption{The one-layer (a) and the slab (b) atomistic models were used in this work to mimic the Fe(221) surface.}
\label{fig:slabs}
\end{figure}

Here, as a guess geometry for the initial and final states, we used the geometries proposed by Zhang et al.\cite{zhang2019interplay} Figure~\ref{fig:slabs}(b) shows the whole slab which consists of 66 iron atoms. By freezing the first two layers of the slab, we obtained the one-layer structure depicted in figure~\ref{fig:slabs}(a) which consists of 22 atoms. Only the nitrogen atoms were allowed to move freely during the structural relaxation, while all the iron atoms were frozen.

\begin{figure}
\centering
\includegraphics[height=6cm]{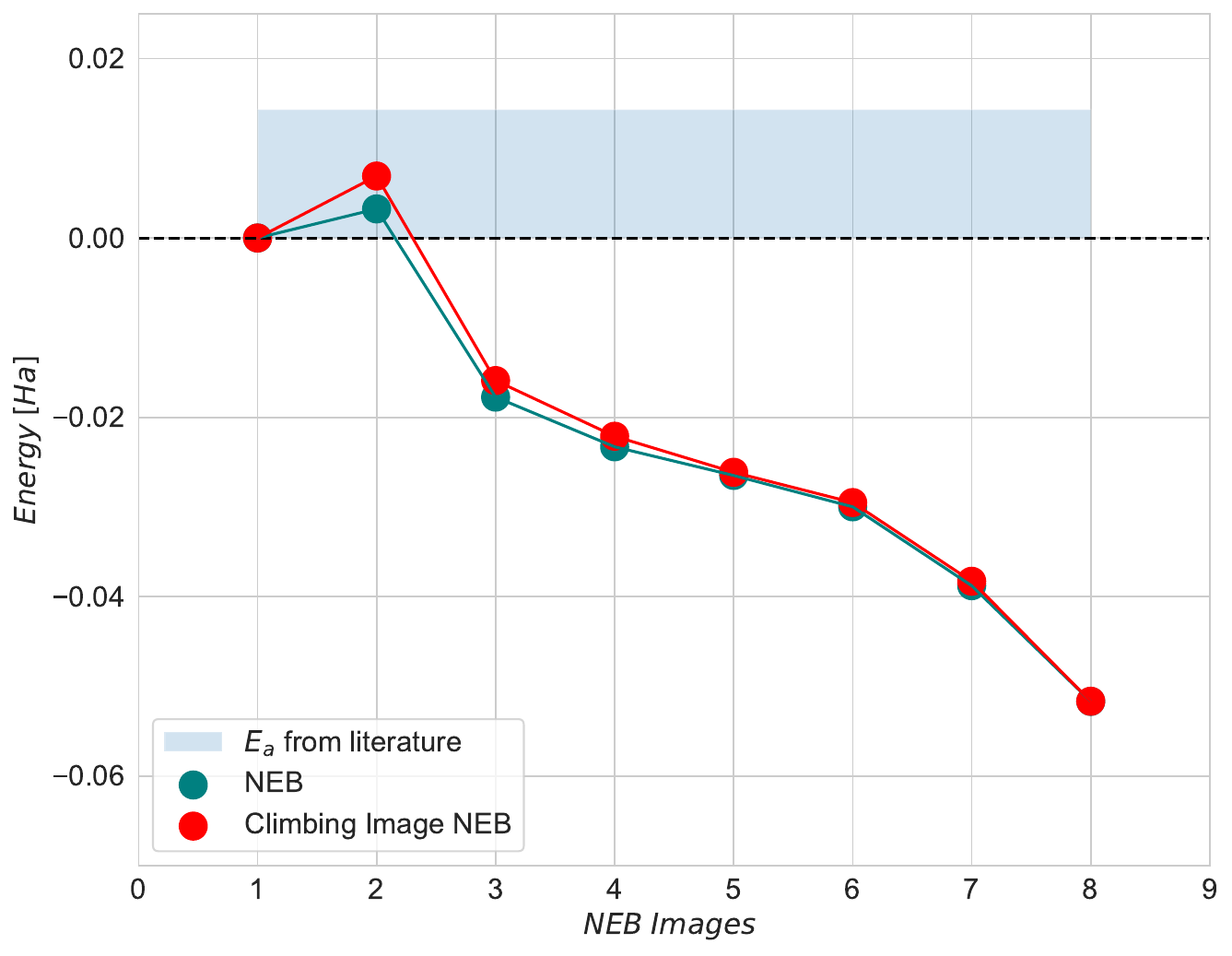}
\caption{NEB simulation of the minimum energy path for the dissociation of N$_2$ molecule on top of the 3-layer Fe(221) model: 8 images of pure NEB (green line/dots) and of Climbing Image refinement (red line/dots). The light blue area represents the range of values for the activation energy available in literature.\cite{zhang2019interplay}}
\label{fig:slab_neb}
\end{figure}

The NEB simulation was carried out on the large Fe(221) surface, and the resulting NEB path was refined further, particularly around the transition state, using the Climbing Image approach~\cite{henkelman2000} to ensure the accurate determination of the Transition State (TS). Results for iron are shown in figure~\ref{fig:slab_neb}. The initial (image 1), transition (image 2), and final (image 8) states from the NEB results were used for all subsequent calculations.

\subsubsection{Choice of the fragment and the active space.}

\begin{figure}
\centering
\includegraphics[height=2.3cm]{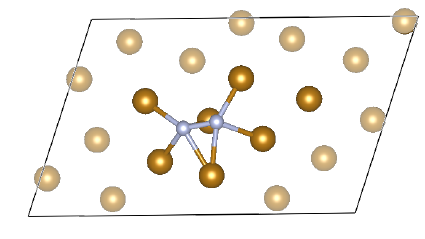}
\caption{The selected sub-system of the single-layer atomistic model (Fe$_{20}$N$_2$). Colour code: gold for Fe and silver for N.}
\label{fig:surface_act}
\end{figure}

For the AVAS active space definition, only a few atoms around the adsorption site were selected from the single-layer atomistic model (Fe$_{20}$N$_2$) as depicted in figure~\ref{fig:surface_act}. The choice of the sub-system (fragment) was based on the observation that the electronic density changes upon adsorption are essentially restricted to the immediate vicinity of the adsorption site, as evidenced by the total density difference plots, see figures \ref{fig:density_first} and \ref{fig:density_end} in Supplementary Information\dag.

\subsubsection{VQE emulator experiments.}

For the single-layer iron surface, we only ran emulator experiments because we observed a minimal difference between the emulator and hardware energy results. The results of the iron cluster experiments confirmed the high performance of Quantinuum's 'H1-1E' noisy emulator backend. Thus, we believe that a similar small difference would have been detected if hardware experiments had been conducted.

\begin{figure}
\centering
\includegraphics[height=6cm]{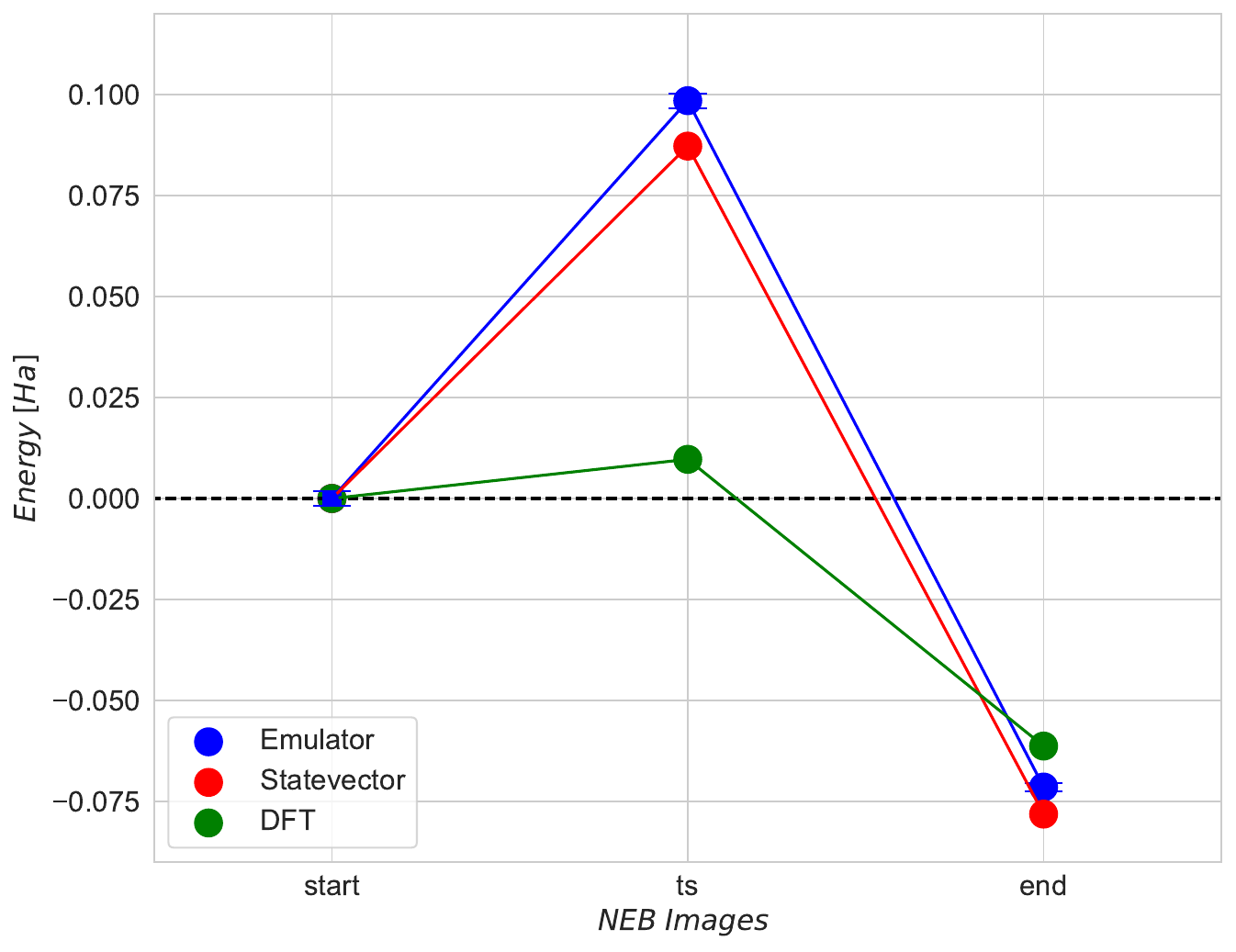}
\caption{Comparison of the activation (E$_a$) and the dissociation (E$_d$) energies by using ADAPT-VQE with a) Quantinuum 'H1-1E' noisy emulator backend, b) Qulacs backend statevector emulator and by using c) DFT. The error bars in the emulator results represent the standard deviation.}
\label{fig:vqe_compare_surf}
\end{figure}

The agreement between the energies computed with the ADAPT-VQE statevector simulator and those from the 'H1-1E' emulator experiments is quite good. The initial geometry has a difference of only 0.74 mHa, as seen in  table~\ref{tab:comp_surf}. By comparing the dissociation energy, it can be seen (figure~\ref{fig:vqe_compare_surf}) that the results from DFT are very consistent to those from ADAPT-VQE for both the statevector and the emulator. In contrast to DFT results, the activation energy is much overestimated.

\begin{table}[h]
\small
  \caption{Comparison of the ADAPT-VQE energy values when running classical statevector and  'H1-1E' emulator calculations}
  \label{tab:comp_surf}
   \begin{tabular*}{0.48\textwidth}{@{\extracolsep{\fill}}llll}
    \hline
    State & State Vector (Ha) & Emulator (Ha) & Deviation (mHa) \\
    \hline
    Initial & -2546.783 & -2546.785 & 1.6\\
    Transition &-2546.696 & -2546.686 & 9.6\\
    Final & -2546.861 & -2546.866 & 5.1\\
    \hline
  \end{tabular*}
\end{table}

\section{Conclusions and future work} \label{conclusions}

We have presented proof-of-concept calculations of the static correlation energies of molecular and periodic models of species involved in the catalytic activation of nitrogen on iron. We have attempted to replicate state-of-the-art results in quantum chemistry by using quantum algorithms on noisy quantum computers with few qubits to demonstrate the current state of quantum computing for chemistry. We focused on 6-qubit computations in the Jordan-Wigner encoding representing 3-orbital active spaces of 2 chemical systems with 3 states each. Efficient circuits were prepared for Hamiltonian averaging by using the ADAPT-VQE ansatz, compiled and processed on backends by using \texttt{pytket}. We applied Post-selection Symmetry Verification error mitigation to reduce the impact of quantum errors. Overall, our results showed an accuracy of about 12 mHa for the best-case cluster systems, which is typical for quantum simulations on current hardware, but about 7 times worse than the so-called ``chemical accuracy" of 1 kcal/mol (1.6 mHa). For the single-layer surface model, the results were much better, as chemical accuracy was achieved for the initial state. These results are very encouraging, even if they do not yet allow us to prove the advantage of quantum solutions for the simulation of complex catalytic systems.

Quantum machinery was not applied to the entire atomistic model - that would be not only expensive but also unnecessary; rather, we used it to describe the immediate vicinity of the reaction site. This is made possible by carefully transforming the mean-field orbitals so that a compact active space is constructed. Crucially, we have found that it is possible to decouple the half-filled Fe-3d band from the Fe-N and N-N bond orbitals, as their mutual correlation turns out to be negligible.

In the case of the single-layer iron surface with 22 atoms, the charge density difference calculations confirmed our hypothesis that we can use only the top iron layer instead of the whole slab and still be able to capture correlation energy equivalent to the CASSCF. As our relative energy results showed, the VQE results were very successful in reproducing the DFT energy trend when studying the activation and dissociation of nitrogen on iron surfaces. Even though this was an approximation, it has clearly shown that large surface models are not necessary and that fragmentation can instead be an accurate way to deal with similar cases.

Future studies of these systems, using both quantum and classical methods, may provide further insight into elementary steps of catalytic ammonia synthesis. A suggestion to better understand the iron catalyst would be to use a simpler facet such as 111 or 211. Even though facets like 221 are more reactive, a simpler facet is an easier problem for today's quantum computers. Other catalysts could be also explored. Examples include ruthenium-based catalysts, which are the most commonly used catalysts for ammonia synthesis after iron, as Ru supported on CaFH can achieve ammonia synthesis at an exceptionally low temperature (50$ ^{\circ}$C), electron-based catalysts such as an ionic O$^{2-}$ compound in which electrons act as the anion, cobalt-based, as it has been found that Co supported on CeO$_2$ or carbon and promoted with Ba has very high activity, nickel-based, which has high activity at low temperatures, and metal nitride catalysts consisting of binary nitride systems based on uranium, cerium, vanadium, molybdenum and rhenium.\cite{morlanes2020development,humphreys2021development,daisley2020comparison,zeinalipour2021comparative}

In addition to investigating other related systems, future work should focus on building up the methodology to reduce errors and increase the size of the calculated systems. It is clear that the Variational Quantum Eigensolver running on noisy quantum processors cannot provide the required accuracy, nor is it applicable to larger problems. A transition to phase estimation algorithms is necessary but requires fault-tolerant quantum computation, which is not yet available. Whether stochastic approximations to phase estimation that can tolerate a certain amount of noise can fill this gap remains to be answered.\cite{Paesani_2017}

While the quantum advantage for Hamiltonian simulation of strongly correlated systems has been postulated \textit{in principle} due to the exponential scaling of classical brute-force algorithms, its \textit{practical} implementation depends on the cost comparison between a classical approximate heuristic algorithm and the quantum method (including the cost of preparing the initial state) for a given use case.\cite{Lee2022} Therefore, follow-up studies could use state-of-the-art approximate classical FCI solvers (such as Stochastic Heat-Bath Configuration Interaction,\cite{SHCI_Umrigar2018} or the Density Matrix Renormalisation Group\cite{DMRG_rev_Reiher2020}) to (1) investigate how complex the electronic states of catalytic species are, (2) determine the scaling of classical approximate ab-initio methods for such systems, and (3) develop prototype end-to-end workflows applicable to large and dense Hamiltonians.

\section*{Acknowledgements}
The authors gratefully acknowledge financial and domain knowledge support from Equinor ASA.
The authors thank Evgeny Plekhanov, Kesha Sorathia and Duncan Gowland for their comments on the manuscript. The authors also thank Isobel Hooper, Brian Neyenhuis and Jenni Strabley for assistance in the hardware experiments. 
\section*{Author contributions}

G.C. directly obtained the results presented in the study. C.D.P. and M.K. assisted with the reaction path calculations. F.E.Z. and A.J. contributed the use case and helped scope the project and conceptualise the study. G.C., C.D.P., M.K. and D.M.R. interpreted the data. M.K. conceptualised the study and supervised the work. G.C. and M.K. wrote and organised the paper. All coauthors participated in joint discussions. 


\balance

\bibliography{rsc}
\bibliographystyle{rsc} 

\clearpage
\onecolumn
\begin{center}
\textbf{\large Electronic Supplementary Information}
\end{center}
\setcounter{equation}{0}
\setcounter{figure}{0}
\setcounter{table}{0}
\setcounter{section}{0}
\setcounter{page}{1}

\section{Atomic positions}

\begin{table}[ht!]
\small
    \caption{Atomic positions of the Fe$_4$N$_2$ for the N$_2$ adsorption}
    \begin{tabular*}{0.48\textwidth}{@{\extracolsep{\fill}}llrrr}
    \hline
    Site n. & Atom &  \multicolumn{3}{l}{Positions (Angstroms)} \\
    \hline                
         1    &      Fe  &  0.2729 & -1.1842 & -0.9998  \\
         2    &      Fe  & -1.3592 &  0.4561 & -0.5239  \\
         3    &      Fe  & -0.9790 & -1.0883 &  1.0747  \\
         4    &      Fe  &  0.3068 &  0.7253 &  1.2245  \\
         5    &      N   &  1.3288 &  0.0599 & -0.1859  \\
         6    &      N   &  0.4298 &  1.0313 & -0.5896  \\  
    \hline
  \end{tabular*}
\end{table}

\begin{table}[ht!]
\small
    \caption{Atomic positions of the Fe$_4$N$_2$ for the N$_2$ activation}
    \begin{tabular*}{0.48\textwidth}{@{\extracolsep{\fill}}llrrr}
    \hline
    Site n. & Atom &  \multicolumn{3}{l}{Positions (Angstroms)} \\
    \hline                
         1    &      Fe  &  0.2958 & -1.0166 & -1.0510  \\
         2    &      Fe  & -1.3984 &  0.5114 & -0.4317  \\
         3    &      Fe  & -0.8682 & -1.1552 &  1.0046  \\
         4    &      Fe  &  0.3232 &  0.8474 &  1.1701  \\
         5    &      N   &  1.2856 & -0.1276 &  0.0849  \\
         6    &      N   &  0.3619 &  0.9407 & -0.7769  \\  
    \hline
  \end{tabular*}
\end{table}

\begin{table}[ht!]
\small
    \caption{Atomic positions of the Fe$_4$N$_2$ for the N$_2$ dissociation}
    \begin{tabular*}{0.48\textwidth}{@{\extracolsep{\fill}}llrrr}
    \hline
    Site n. & Atom &  \multicolumn{3}{l}{Positions (Angstroms)} \\
    \hline                
         1    &      Fe  &  0.4544 & -0.9750 & -1.0525  \\
         2    &      Fe  & -1.4660 &  0.5748 & -0.5102  \\
         3    &      Fe  & -0.8155 & -1.1256 &  0.9120  \\
         4    &      Fe  &  0.5954 &  0.8987 &  0.8914  \\
         5    &      N   &  1.2257 & -0.6628 &  0.4615  \\
         6    &      N   &  0.0061 &  1.2899 & -0.7021  \\    
    \hline
  \end{tabular*}
\end{table}

\begin{table}[ht!]
\small
    \caption{Atomic positions of the Fe$_3$N$_2$ for the N$_2$ adsorption}
    \begin{tabular*}{0.48\textwidth}{@{\extracolsep{\fill}}llrrr}
    \hline
    Site n. & Atom &  \multicolumn{3}{l}{Positions (Angstroms)} \\
    \hline                
         1    &      Fe  &  -1.2363 & -0.5717 &  0.0002 \\
         2    &      Fe  &   0.8587 & -0.5618 &  1.0672 \\
         3    &      Fe  &   0.8585 & -0.5622 & -1.0672 \\
         4    &      N   &  -0.2406 &  0.8477 & -0.7215 \\
         5    &      N   &  -0.2403 &  0.8480 &  0.7213 \\  
    \hline
  \end{tabular*}
\end{table}

\begin{table}[ht!]
\small
    \caption{Atomic positions of the Fe$_3$N$_2$ for the N$_2$ activation}
    \begin{tabular*}{0.48\textwidth}{@{\extracolsep{\fill}}llrrr}
    \hline
    Site n. & Atom &  \multicolumn{3}{l}{Positions (Angstroms)} \\
    \hline                
         1    &      Fe  &  -1.2417 & -0.4712 & -0.0332  \\
         2    &      Fe  &   0.9278 & -0.5001 &  1.0778  \\
         3    &      Fe  &   0.8116 & -0.5615 & -1.0570  \\
         4    &      N   &  -0.2095 &  0.7987 & -0.9081  \\
         5    &      N   &  -0.2881 &  0.7341 &  0.9205  \\  
    \hline
  \end{tabular*}
\end{table}

\begin{table}[ht!]
\small
    \caption{Atomic positions of the Fe$_3$N$_2$ for the N$_2$ dissociation}
    \begin{tabular*}{0.48\textwidth}{@{\extracolsep{\fill}}llrrr}
    \hline
    Site n. & Atom &  \multicolumn{3}{l}{Positions (Angstroms)} \\
    \hline                
         1    &      Fe  &  -1.17445 & -0.12095 &  0.00018  \\
         2    &      Fe  &   0.92744 & -0.48562 &  1.14121  \\
         3    &      Fe  &   0.92714 & -0.48608 & -1.14127  \\
         4    &      N   &  -0.34024 &  0.54605 & -1.35633  \\
         5    &      N   &  -0.33988 &  0.54659 &  1.35619  \\ 
    \hline
  \end{tabular*}
\end{table}

\begin{table}[ht!]
\small
    \caption{Atomic positions of the single-layer surface Fe$_2$N$_2$ for the N$_2$ adsorption}
    \begin{tabular*}{0.48\textwidth}{@{\extracolsep{\fill}}lllll}
    \hline
    Site n. & Atom &  \multicolumn{3}{l}{Positions (crystal coordinates)} \\
    \hline                
         1    &      Fe  &  0.5231 & 0.4625 & 0.5375  \\
         2    &      Fe  &  0.5257 & 0.9625 & 0.5401  \\
         3    &      Fe  &  0.0257 & 0.9630 & 0.5407  \\
         4    &      Fe  &  0.0261 & 0.4638 & 0.5412  \\
         5    &      Fe  &  0.2203 & 0.8513 & 0.5557  \\
         6    &      Fe  &  0.2151 & 0.3500 & 0.5554  \\
         7    &      Fe  &  0.7277 & 0.3507 & 0.5560  \\
         8    &      Fe  &  0.7191 & 0.8503 & 0.5561  \\
         9    &      Fe  &  0.4159 & 0.7459 & 0.5806  \\
        10    &      Fe  &  0.9164 & 0.2427 & 0.5795  \\
        11    &      Fe  &  0.9202 & 0.7453 & 0.5806  \\
        12    &      Fe  &  0.4186 & 0.2460 & 0.5819  \\
        13    &      Fe  &  0.1105 & 0.6350 & 0.6007  \\
        14    &      Fe  &  0.6169 & 0.6434 & 0.5973  \\
        15    &      Fe  &  0.6133 & 0.1358 & 0.5989  \\
        16    &      Fe  &  0.1153 & 0.1387 & 0.5996  \\
        17    &      Fe  &  0.3083 & 0.5176 & 0.6169  \\
        18    &      Fe  &  0.8069 & 0.0207 & 0.6155  \\
        19    &      Fe  &  0.8011 & 0.5223 & 0.6185  \\
        20    &      Fe  &  0.3035 & 0.0212 & 0.6157  \\
        21    &      N   &  0.5589 & 0.3908 & 0.6115  \\
        22    &      N   &  0.5335 & 0.4842 & 0.6275  \\
    \hline
  \end{tabular*}
\end{table}

\begin{table}[ht!]
\small
    \caption{Atomic positions of the single-layer surface Fe$_2$N$_2$ for the N$_2$ activation}
    \begin{tabular*}{0.48\textwidth}{@{\extracolsep{\fill}}lllll}
    \hline
    Site n. & Atom &  \multicolumn{3}{l}{Positions (crystal coordinates)} \\
    \hline                 
         1    &      Fe  &  0.5231 & 0.4625 & 0.5375  \\
         2    &      Fe  &  0.5257 & 0.9625 & 0.5401  \\
         3    &      Fe  &  0.0257 & 0.9630 & 0.5407  \\
         4    &      Fe  &  0.0261 & 0.4638 & 0.5412  \\
         5    &      Fe  &  0.2203 & 0.8513 & 0.5557  \\
         6    &      Fe  &  0.2151 & 0.3500 & 0.5554  \\
         7    &      Fe  &  0.7277 & 0.3507 & 0.5560  \\
         8    &      Fe  &  0.7191 & 0.8503 & 0.5561  \\
         9    &      Fe  &  0.4159 & 0.7459 & 0.5806  \\
        10    &      Fe  &  0.9164 & 0.2427 & 0.5795  \\
        11    &      Fe  &  0.9202 & 0.7453 & 0.5806  \\
        12    &      Fe  &  0.4186 & 0.2460 & 0.5819  \\
        13    &      Fe  &  0.1105 & 0.6350 & 0.6007  \\
        14    &      Fe  &  0.6169 & 0.6434 & 0.5973  \\
        15    &      Fe  &  0.6133 & 0.1358 & 0.5989  \\
        16    &      Fe  &  0.1153 & 0.1387 & 0.5996  \\
        17    &      Fe  &  0.3083 & 0.5176 & 0.6169  \\
        18    &      Fe  &  0.8069 & 0.0207 & 0.6155  \\
        19    &      Fe  &  0.8011 & 0.5223 & 0.6185  \\
        20    &      Fe  &  0.3035 & 0.0212 & 0.6157  \\
        21    &      N   &  0.5688 & 0.3879 & 0.6059   \\
        22    &      N   &  0.5300 & 0.4985 & 0.6242  \\
    \hline
  \end{tabular*}
\end{table}

\begin{table}[ht!]
\small
    \caption{Atomic positions of the single-layer surface Fe$_2$N$_2$ for the N$_2$ dissociation}
    \begin{tabular*}{0.48\textwidth}{@{\extracolsep{\fill}}lllll}
    \hline
    Site n. & Atom &  \multicolumn{3}{l}{Positions (crystal coordinates)} \\
    \hline                       
         1    &      Fe  &  0.5231 & 0.4625 & 0.5375  \\
         2    &      Fe  &  0.5257 & 0.9625 & 0.5401  \\
         3    &      Fe  &  0.0257 & 0.9630 & 0.5407  \\
         4    &      Fe  &  0.0261 & 0.4638 & 0.5412  \\
         5    &      Fe  &  0.2203 & 0.8513 & 0.5557  \\
         6    &      Fe  &  0.2151 & 0.3500 & 0.5554  \\
         7    &      Fe  &  0.7277 & 0.3507 & 0.5560  \\
         8    &      Fe  &  0.7191 & 0.8503 & 0.5561  \\
         9    &      Fe  &  0.4159 & 0.7459 & 0.5806  \\
        10    &      Fe  &  0.9164 & 0.2427 & 0.5795  \\
        11    &      Fe  &  0.9202 & 0.7453 & 0.5806  \\
        12    &      Fe  &  0.4186 & 0.2460 & 0.5819  \\
        13    &      Fe  &  0.1105 & 0.6350 & 0.6007  \\
        14    &      Fe  &  0.6169 & 0.6434 & 0.5973  \\
        15    &      Fe  &  0.6133 & 0.1358 & 0.5989  \\
        16    &      Fe  &  0.1153 & 0.1387 & 0.5996  \\
        17    &      Fe  &  0.3083 & 0.5176 & 0.6169  \\
        18    &      Fe  &  0.8069 & 0.0207 & 0.6155  \\
        19    &      Fe  &  0.8011 & 0.5223 & 0.6185  \\
        20    &      Fe  &  0.3035 & 0.0212 & 0.6157  \\
        21    &      N   &  0.5815 & 0.3873 & 0.6007  \\
        22    &      N   &  0.4396 & 0.6527 & 0.6380  \\
    \hline
  \end{tabular*}
\end{table}

\section{The Fe$_3$N$_2$ cluster}

\begin{figure}[ht!]
 \centering
 \includegraphics[height=2cm]{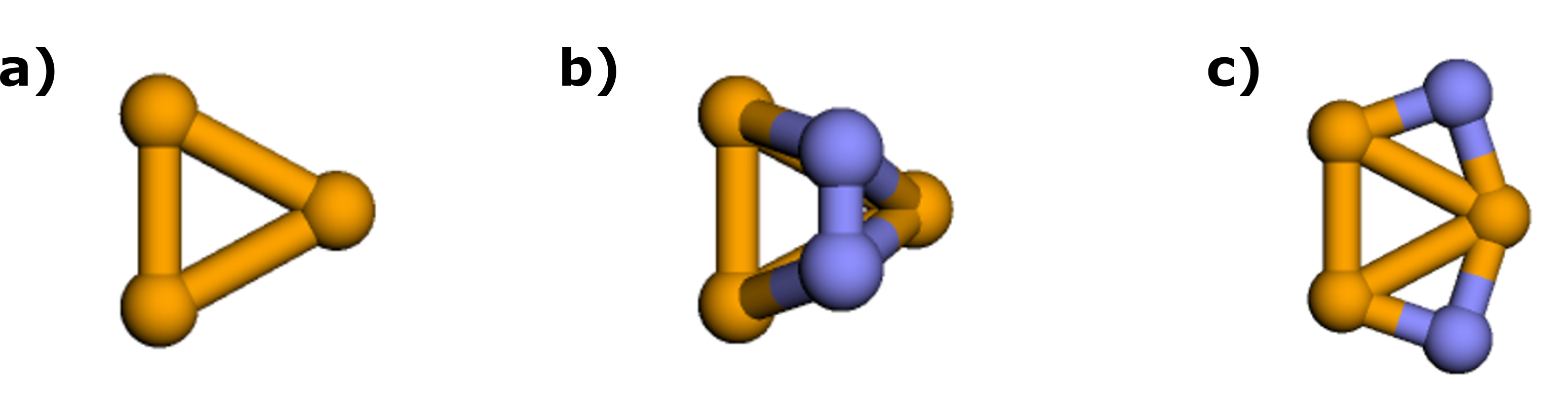}
 \caption{Visualisation of the optimised geometries of the a)Fe$_3$ cluster b) N$_2$ adsorption on Fe$_3$ and c) N$_2$ dissociation on Fe$_3$ using InQuanto-NGLView. Colour code: orange for Fe and blue for N.}
 \label{fig:clusterFe3}
\end{figure}

Structural relaxation with spin-polarised PBE/PAW DFT using Quantum Espresso and PBE0 using PySCF were performed for the Fe$_3$ cluster. To build the model cluster systems, we first optimised the Fe$_3$ molecule in an equilateral triangular geometry (figure~\ref{fig:clusterFe3}(a)) with the net spin of S$=$4 (2.67 $\mu$B per atom), 2S+1$=$9 and the bond length of 2.04 \AA \ as the lowest energy state for Fe$_3$.\cite{pakiari2010detailed,liu2018heterogeneous,gutsev2021evolution} From this structure, all relevant products were derived by adding two nitrogen atoms and re-optimising (initial geometry, figure~\ref{fig:clusterFe3}(b)) or only by stretching the nitrogen atoms and re-optimising to obtain the final geometry (figure~\ref{fig:clusterFe3}(c)). Nudged Elastic Band method (NEB) was also employed to find the minimum energy reaction path and transition state for Fe$_3$N$_2$ with eight NEB images.

\begin{figure}[ht!]
 \centering
\includegraphics[height=6cm]{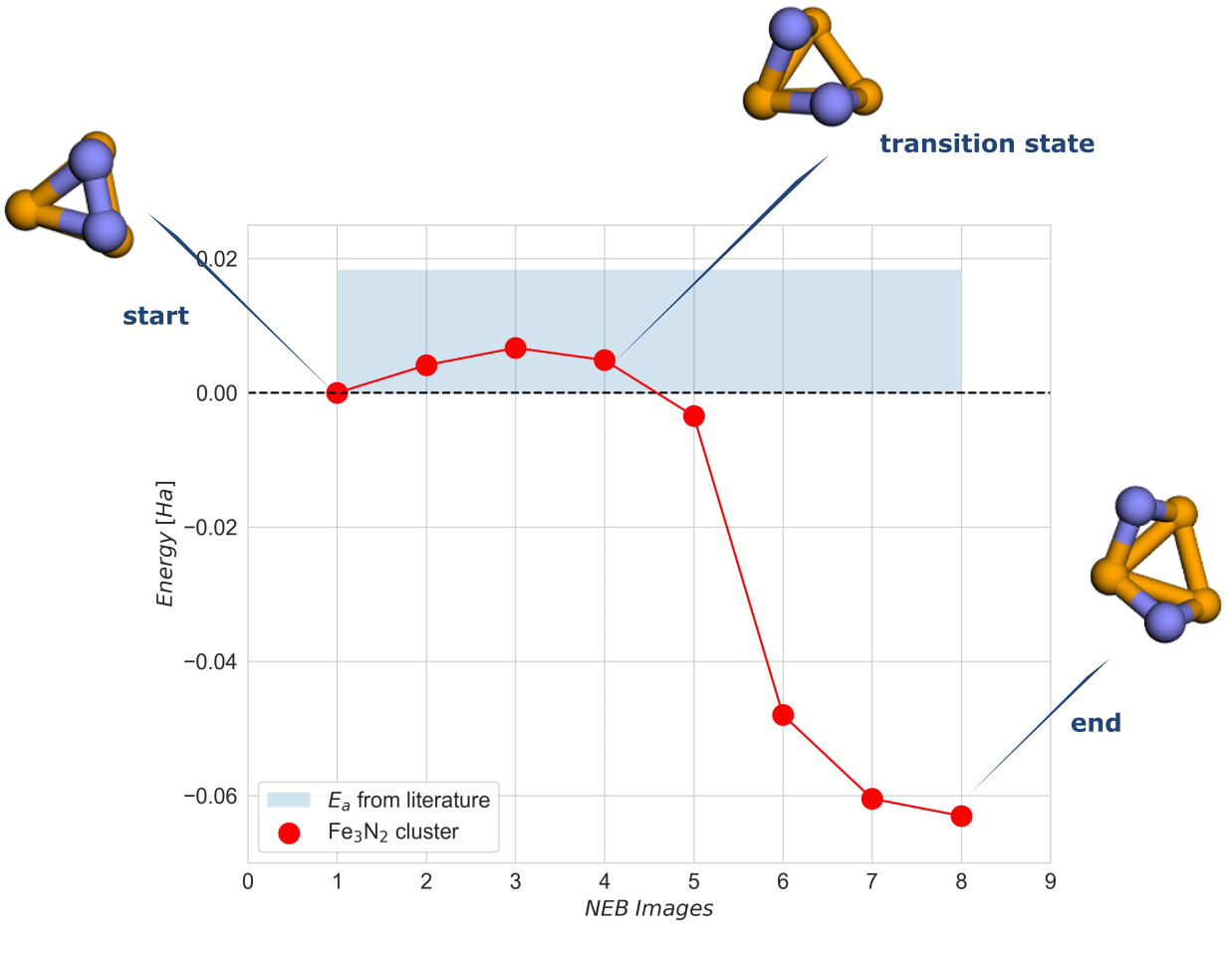}
 \caption{ NEB simulation of the minimum energy path for the dissociation of the N$_2$ molecule on the Fe$_3$N$_2$ cluster. The light blue region represents the range of activation energy values available in the literature.\cite{gutsev2021evolution,liu2018heterogeneous} Colour code: orange for Fe and blue for N.}
\end{figure}

\section{Active space selection}

The nature of the CAS-CI wavefunctions depends strongly on the choice of active orbitals. When only the d manifold of the Fe atoms is active, the resulting state is an equal mixture of 6 configurations that differ in the placement of electrons on the d orbitals. As shown in table~\ref{tbl:fe3_NFe}, adding the orbitals of the nitrogen atoms results in almost zero contribution of the excited determinants in the wavefunction.

\begin{table}[ht!]
\small
    \caption{Largest CI components of the transition state for the Fe$_3$N$_2$ cluster when the 2p orbitals of the two nitrogen atoms and 3d orbitals of the four irons are selected}
    \label{tbl:fe3_NFe}
    \begin{tabular*}{0.4\textwidth}{@{\extracolsep{\fill}}lll}
    \hline
    alpha occ-orbitals & beta occ-orbitals & CI coefficient \\
    \hline
    $[0 1 2]$ & $[0 1 2]$ &  0.98873  \\   
    $[0 1 3]$ & $[0 1 3]$ & -0.04063   \\
    $[0 1 3]$ & $[0 1 4]$ &  0.03449   \\
    $[0 1 4]$ & $[0 1 3]$ &  0.03449   \\ 
    $[0 1 4]$ & $[0 1 4]$ & -0.05175   \\
    \hline
  \end{tabular*}
\end{table}

\begin{figure}[ht!]
\centering
\includegraphics[height=1.8cm]{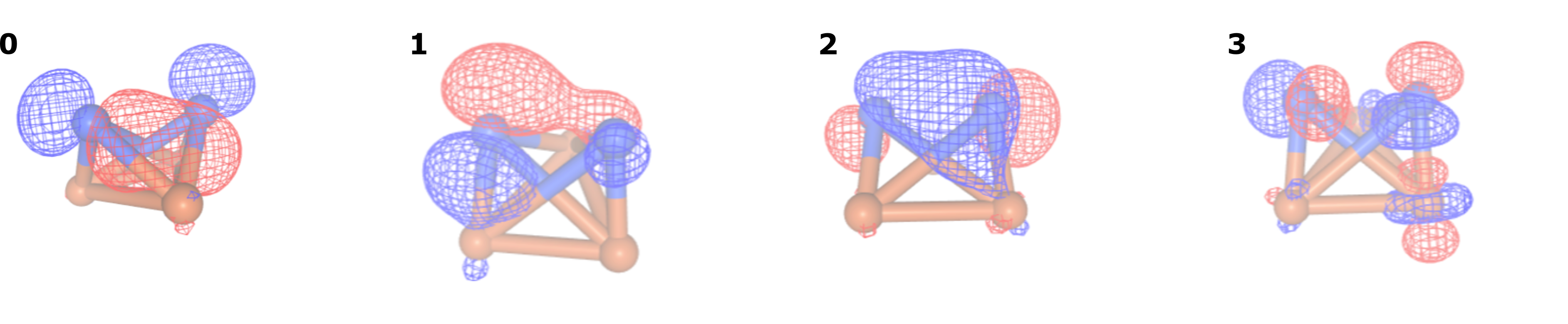}
\caption{Schematic representation of the main orbitals presented in table ~\ref{tbl:fe3_NFe} for the Fe$_3$N$_2$ cluster.}
\label{fig:orbitals_Fe3N2}
\end{figure}

\begin{figure}[ht!]
\centering
\includegraphics[height=1.6cm]{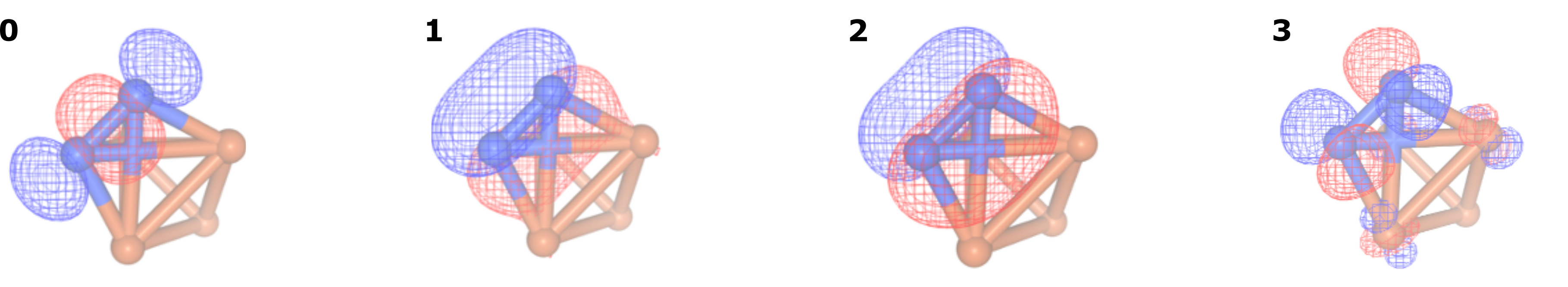}
\caption{Schematic representation of the main orbitals presented in table ~\ref{tbl:fe4_NFe} for the Fe$_4$N$_2$ cluster.}
\label{fig:orbitals_Fe4N2}
\end{figure}

\section{DFT density difference}

\begin{figure}[ht!]
\centering
\includegraphics[height=5cm]{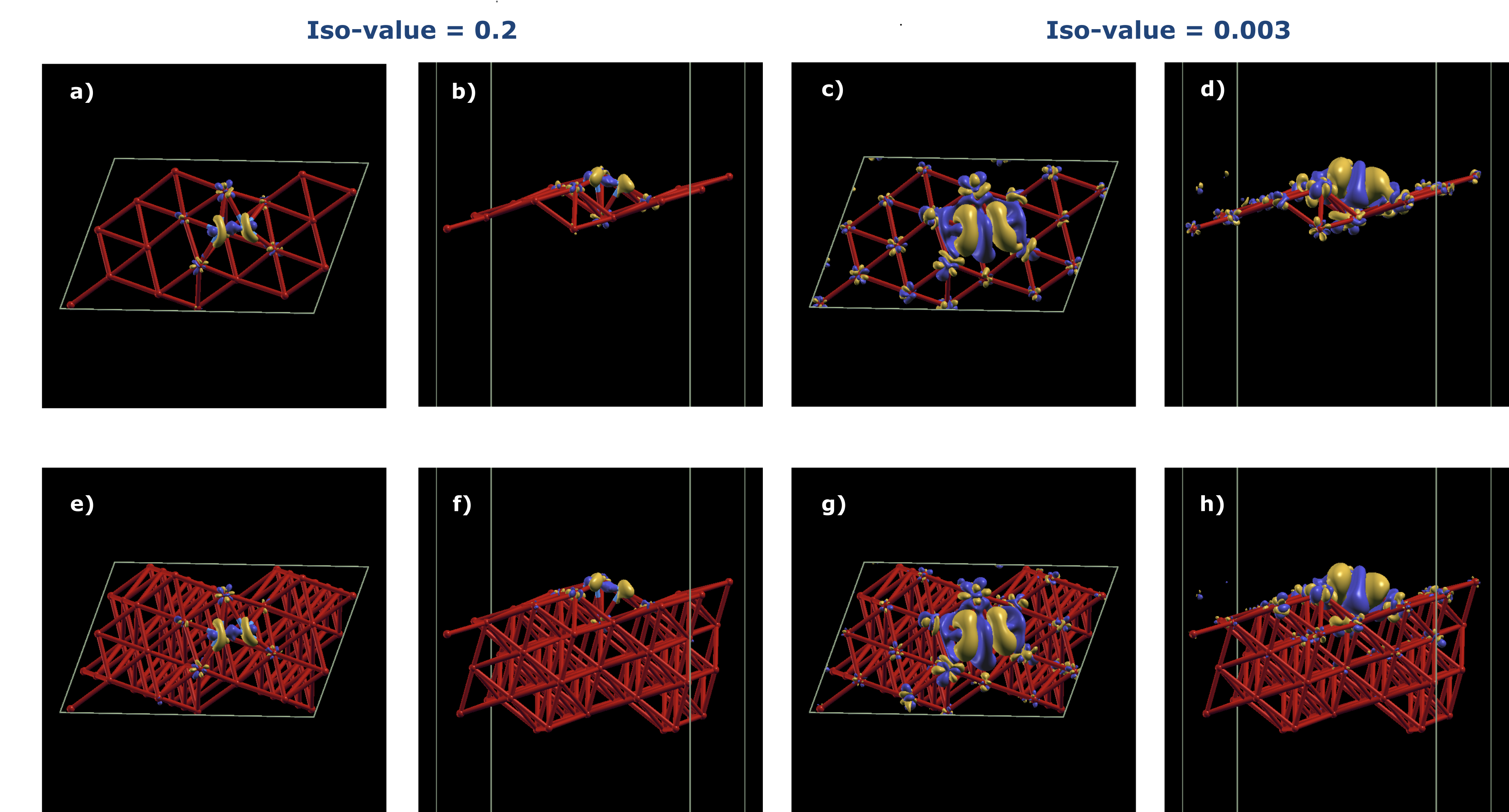}
\caption{DFT density difference: $\Delta \rho = \rho _{Fe/NN}-(\rho _{Fe}+\rho _{NN})$, for the initial geometry obtained at two different iso-values (in units of electron/bohr$^3$). Top (a and c) and side (b and d) views of the single-layer surface model with 22 atoms and top (e and g) and side (f and h) views of the iron slab with 66 atoms. Blue and yellow volumes represent loss or gain in electron density, respectively, when the adsorbate+substrate is formed from the isolated constituents.}
\label{fig:density_first}
\end{figure}

\begin{figure}[ht!]
\centering
\includegraphics[height=5cm]{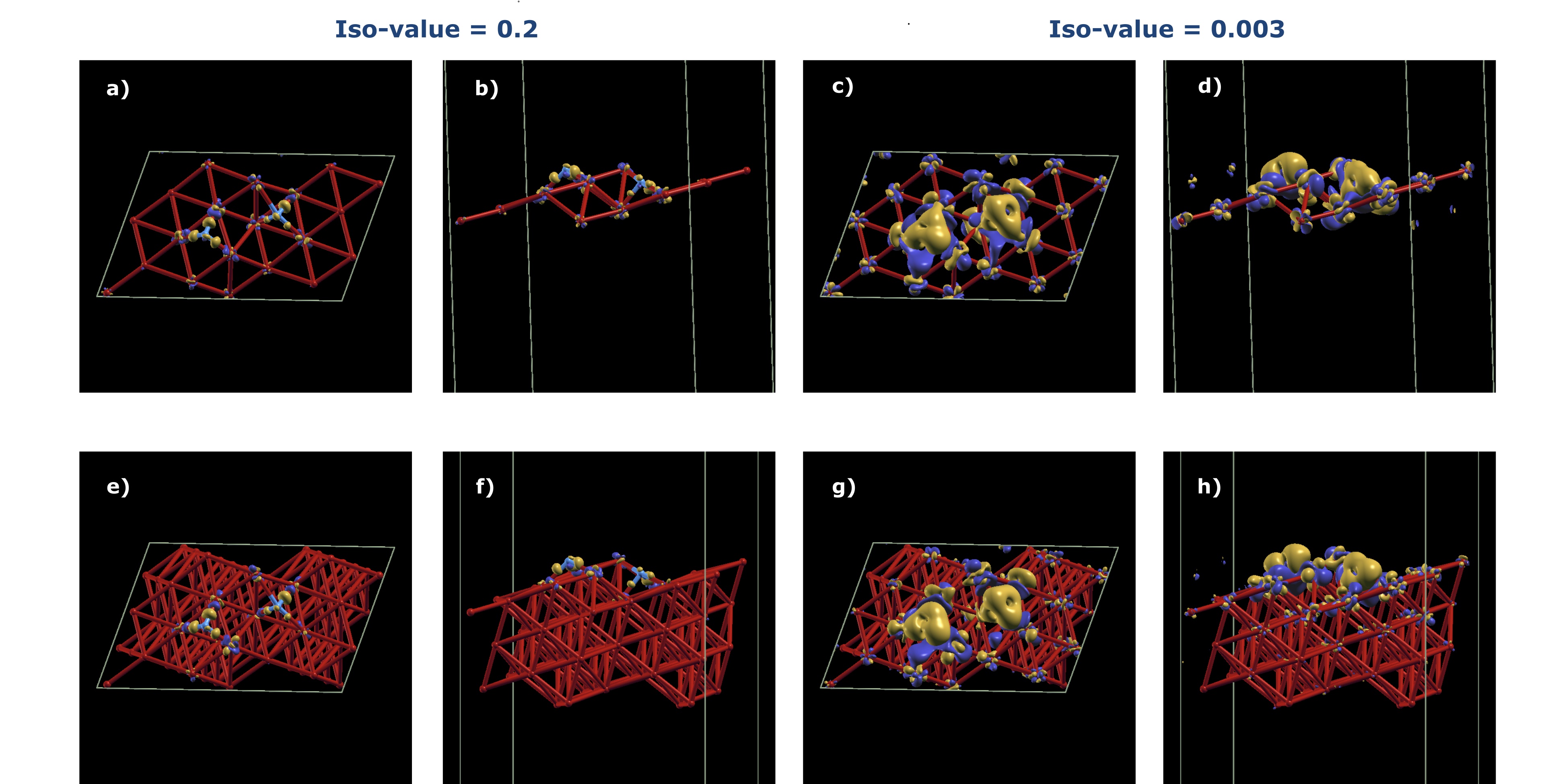}
\caption{DFT density difference: $\Delta \rho = \rho _{Fe/NN}-(\rho _{Fe}+\rho _{NN})$, for the final geometry obtained at two different iso-values (in units of electron/bohr$^3$). Top (a and c) and side (b and d) views of the single-layer surface model with 22 atoms and top (e and g) and side (f and h) views of the iron slab with 66 atoms. Blue and yellow volumes represent loss or gain in electron density, respectively, when the adsorbate+substrate is formed from the isolated constituents.}
\label{fig:density_end}
\end{figure}

The charge density difference after the N$_2$ dissociation was calculated and plotted using Quantum ESPRESSO. As shown in figure\ref{fig:density_first} (initial state) and in figure\ref{fig:density_end} (final state). These density difference results for different iso-values revealed an interesting result as the iron atoms involved in these correlation interactions are predominantly located in the topmost layer. This is an important finding for further approximations so that we can simplify our slab model and just use a single-layer surface model instead.

\section{Simple noise-free and stochastic simulations}

The evaluation of a quantum measurement can be carried out directly by statevector methods, which involve performing linear algebra of the gate operations directly on an explicit 2$^N$ dimensional representation of the quantum state. This method directly returns the statevector of the system, neglecting the stochastic nature of quantum measurements.

\begin{figure}[ht!]
\centering
    \includegraphics[height=5cm]{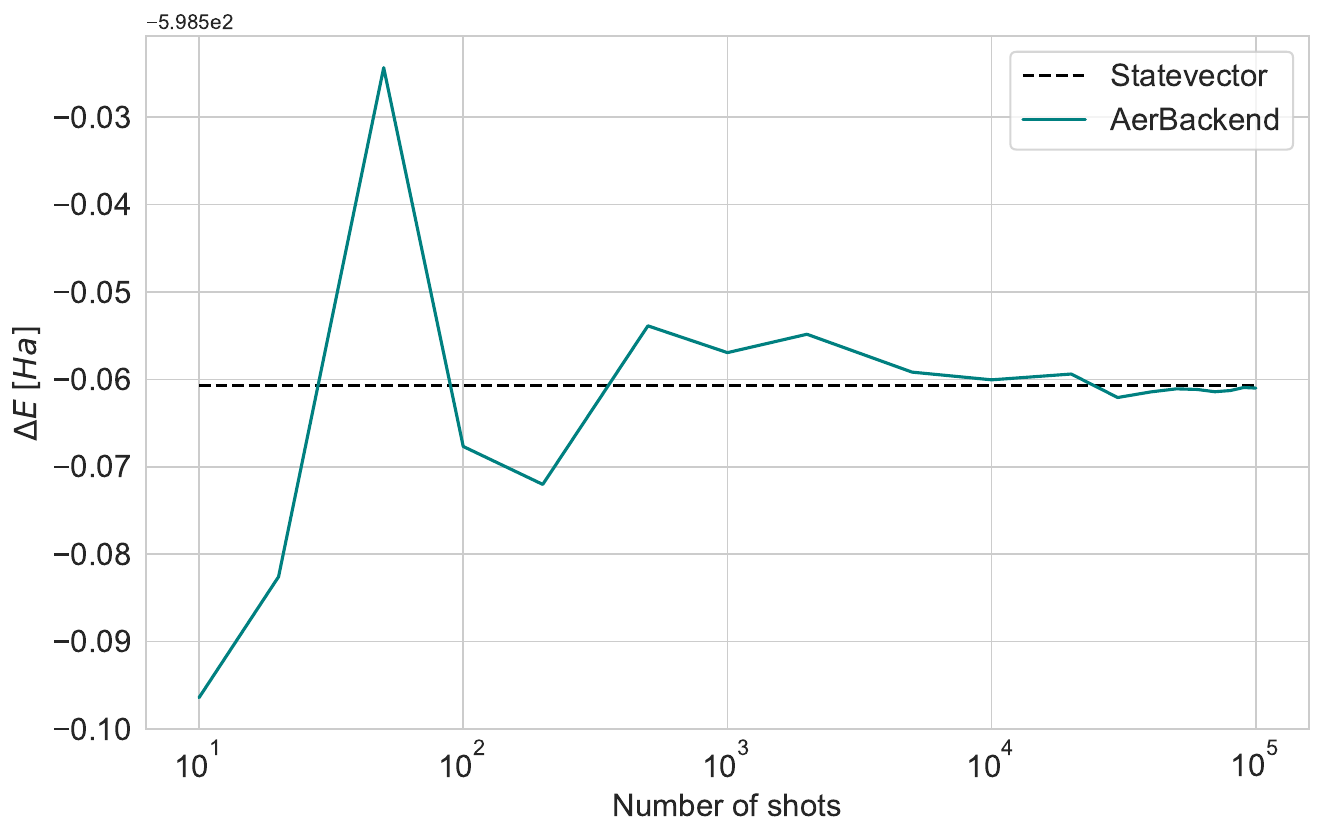}
    \caption{Expectation value obtained by running a noise-free simulation (statevector) and a stochastic simulation (AerBackend). The Fe$_3$N$_2$ cluster's initial geometry was consistently utilized across all simulations.}
    \label{fig:statevector}
\end{figure}

Next, we performed the simulation with Aer, the noiseless stochastic quantum simulator being part of Qiskit \cite{Qiskit}. This simulation does not have any quantum errors, such as the decoherence of the quantum state during processing. However, it does incorporate the stochastic nature of measuring the system, which results in a probabilistic collapse into a computational basis state at the end of the simulation. Each individual shot will return a measurement outcome of +1 or -1. By conducting multiple shots, one can estimate the expectation value of each Pauli string in the Hamiltonian. Increasing the number of shots will lead to higher precision in the estimation of each expectation value. This results in increased precision when combined to determine the expectation value of the full Hamiltonian. Figure~\ref{fig:statevector} shows the convergence of the energy with the number of shots. This shows that at approximately 1000 shots, the system averaging is sufficiently accurate to reasonably recreate the statevector result (within a margin of 0.01 Ha). 

\begin{figure}[ht!]
    \centering
\includegraphics[height=5cm]{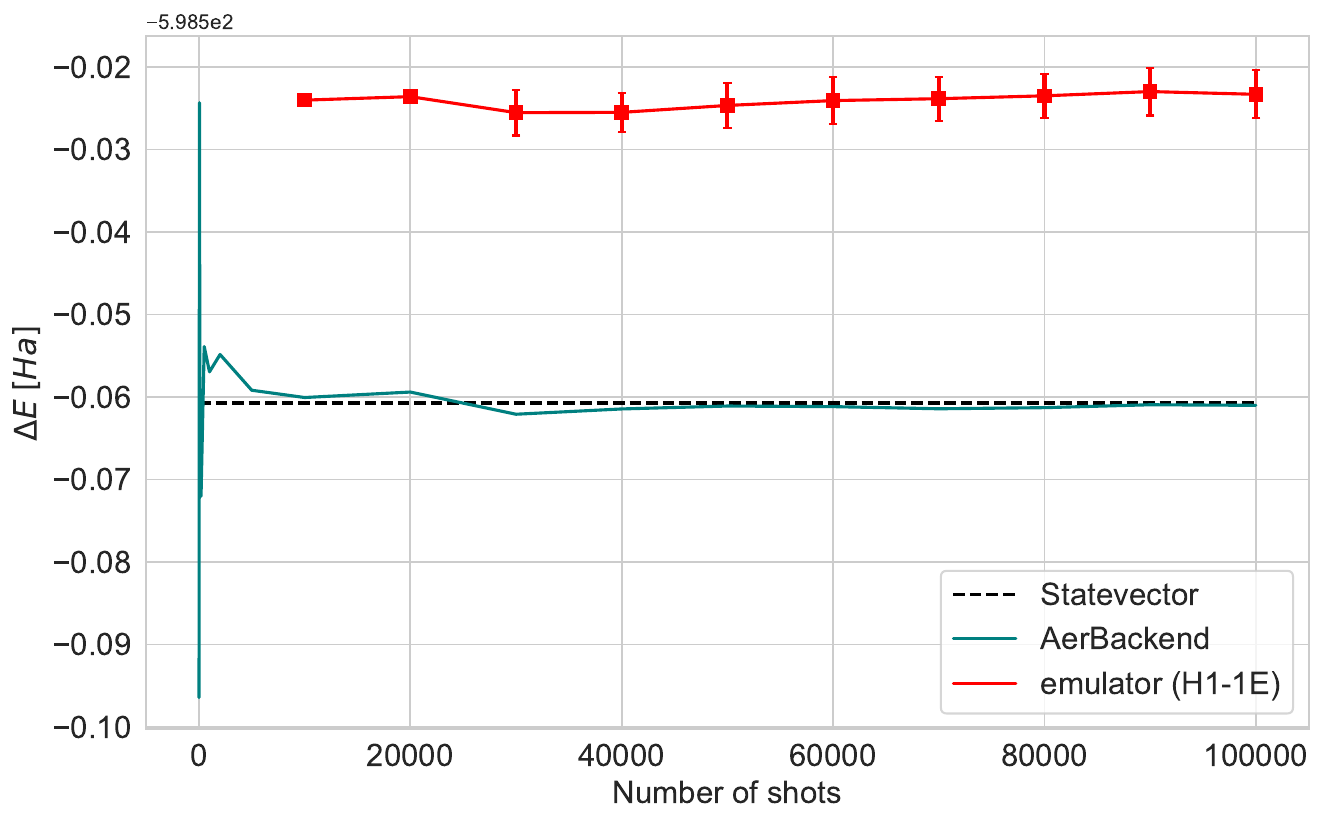}
    \caption{Expectation value obtained by a noise-free simulation (Statevector), a stochastic simulation (AerBackend) and a calculation with the Quantinuum 'H1-1E' noisy emulator backend (10K shots). The Fe$_3$N$_2$ cluster's initial geometry was consistently utilized across all simulations. The error bars in the results represent the standard deviation.}
    \label{fig:comparison}
\end{figure}

Having established the convergence of the expectation value of energy with respect to number of measurements in the absence of noise, we proceeded to experiments with the 'H1-1E' quantum emulator, the `digital twin' of the H1-1 quantum computer. As expected, Figure~\ref{fig:comparison} shows that the emulator's noise model leads to an increase in energy and poorer accuracy.  

\end{document}